\documentclass[structabstract]{aa}  
\usepackage{graphicx,natbib}
\bibpunct{(}{)}{;}{a}{}{,}
\usepackage{txfonts}
\usepackage{deluxetable} 

\newcommand\msol{\ensuremath{\,\mbox{\it M}_{\odot}}}

\newcommand\lta{\mathrel{\hbox{\raise 0.6 ex \hbox{$<$}\kern
                   -1.8 ex\lower .5 ex\hbox{$\sim$}}}}
\newcommand\gta{\mathrel{\hbox{\raise 0.6 ex \hbox{$>$}\kern
                   -1.7 ex\lower .5 ex\hbox{$\sim$}}}}
 
\newcommand{\scrbox}[1]{\ensuremath{{\mbox{\scriptsize #1}}}}

\newcommand{\teff}{{\ensuremath{T_{\scrbox{eff}}}}}
\newcommand{\Msol}{\ensuremath{\,\mbox{\it M}_{\odot}}}
\newcommand{\Mstar}{\ensuremath{\it \,M_{*}}}

\newcommand{\Dturb}{\ensuremath{D_{\scrbox{T}}}}

\newcommand{\MS}{main--sequence}
\newcommand{\gr}{\ensuremath{g_{\scrbox{rad}}}}

\newcommand{\DM}{\ensuremath{ \log \Delta M/M_{*}}}
\newcommand{\DMsol}{\ensuremath{ \log \Delta M/M_{\odot}}}

\renewcommand{\H}{\mbox{H}}
\newcommand{\He}{\mbox{He}}

\newcommand{\Fe}{\mbox{Fe}}

\newcommand{\Cr}{\mbox{Cr}}
\newcommand{\Ca}{\mbox{Ca}}
\newcommand{\Ni}{\mbox{Ni}}

\begin{document}
   \title{Horizontal Branch evolution, metallicity and sdB stars}

   \author{G. Michaud \inst{1,2} \and J. Richer
          \inst{2}
          \and
          O. Richard\inst{3}
          }

   \institute{LUTH, Observatoire de Paris, CNRS, Universit\'e Paris Diderot,
     5 Place Jules Janssen,
     92190 Meudon, FRANCE
     \and  
   D\'epartement de Physique, Universit\'e de Montr\'eal,
       Montr\'eal, PQ, H3C~3J7, CANADA\\
              \email{michaudg@astro.umontreal.ca,jacques.richer@umontreal.ca}
         \and
             Universit\'e Montpellier II - GRAAL, CNRS - UMR 5024, place Eug\`ene Bataillon, 
34095 Montpellier, FRANCE\\
             \email{Olivier.Richard@univ-montp2.fr}
                  }

   \date{\today}


  \abstract
   {Abundance anomalies have  been observed  in field sdB stars and in nearly all Horizontal Branch (HB) stars of globular clusters with $\teff >
11\,000$\,K   whatever be the  cluster metallicity.}
   {The aim is to determine the  abundance variations to be expected in sdB stars and in HB stars of metallicities  $Z \ge 10^{-4}$ 
   and what observed abundances teach us about hydrodynamical processes competing with atomic diffusion. }
   {Complete stellar evolution models, including the effects of atomic diffusion and radiative acceleration, have been computed from the zero age \MS{} for metallicities of $Z_0= 0.0001$, 0.001, 0.004 and 0.02.  On the HB the masses were selected to cover the \teff{} interval from 7000 to 37000\,K. Some 60 evolutionary HB models were calculated.  The calculations of surface abundance anomalies during the horizontal branch depend on one parameter, the surface mixed mass.
   }
   {For sdB stars with $\teff <37000$\,K and for  HB stars with $\teff > 11\,000$\,K  in all observed clusters, independent of metallicity, it was found that most observed abundance
   anomalies  (even up to $\sim \times$\,200) were compatible, within error bars, with expected abundances. 
   A mixed mass of $\sim 10^{-7} \msol$ was determined by comparison with observations.}
   {Observations of globular cluster HB stars with $\teff > 11\,000$\,K  and of sdB stars with $\teff < 37\,000$\,K   suggest that most observed abundance anomalies can be explained by element separation driven by radiative acceleration occuring at a mass fraction of $\sim 10^{-7} \msol$. Mass loss or turbulence appear to limit the separation between $ 10^{-7} \msol$ and the surface.}

   \keywords{Horizontal Branch -- sdB stars -- Stellar evolution -- chemical composition
               }

   \maketitle
%

\section{Astrophysical context}

Large abundance anomalies have been observed on the horizontal branch (HB) of NGC6752, NGC1904, NGC2808, M15 and M13 
 \citep{BehrCoMcetal99,BehrCoMc2000,Behr2003,MoehlerSwLaetal2000,FabbianReGretal2005,PaceRePietal2006}: whereas those stars 
 cooler than about 11000\,K have the same composition as giants,
those hotter than 11000\,K usually have larger abundances of some metals by large factors. This occurs in all clusters having  blue enough HB  stars irrespective of their metallicity as defined by their giant branch stars.

Field sdOB stars are observed to have large abundance anomalies compared to Pop I stars (for a review see \citealt{Heber2009}).  They were already recognized by \citet{SargentSe68} to have a surface composition different from the one with which they formed.  While cool ($\teff < 10^{4}$\,K) field Pop II stars have low $Z$, the sdBs\footnote{We use the expression HB stars for Horizontal Branch stars in clusters, and sdB stars for those in the field.  In practice, since calculations are always done for a given metallicity, they are always done for cluster stars causing  apparent contradictions to that rule when comparisons are made to field stars.} which must come from the red giants often have iron peak abundances which are solar or even larger. While sdBs correspond to the blue end of the HB, the hotter sdO stars 
are apparently a mixed bag of post HB stars and other highly evolved evolutionary stages.  Spectroscopically, sdOs have much more diverse characteristics than sdBs.  In this paper we extend the $\teff$ coverage of evolutionary models to sdBs but not to sdOs.  

In preceding papers \citep{MichaudRiRi2007,MichaudRiRi2008} the evolution of a Pop II star\footnote{In this series of papers, the relative values of the $\alpha$ elements are increased following \citet{VandenBergSwRoetal2000}.  Models are labeled according to their original $Z_0$ value calculated before the $\alpha$ correction. See also Table 1 of \citealt{RichardMiRietal2002}. This correction was not applied in the $Z_0 = 0.02$ models.}  with $Z_0 = 10^{-4}$
   was followed from the zero age \MS{} to the middle of the HB and comparisons were made to observed abundance anomalies in M\,15. The overabundances are explained by atomic diffusion driven by radiative  accelerations in stars with $\teff > 11000$\,K and the sudden break in anomalies at 11000\,K was shown to be related to observed rotation velocities \citep{QuievyChMietal2009}.  Given the relatively large observational error bars, the anomalies appeared compatible with 
   a simple diffusion model involving only one parameter, the mass of the outer region mixed  by turbulence.  It was 
   determined by \citet{MichaudRiRi2008} to be about 10$^{-7}$\,\Mstar.
   
Extending these calculations to higher metallicities allows comparing to field sdOB stars.  While these have the disadvantage that their original metallicity is unknown, some of them are much closer than any globular cluster and so their surface composition can be determined more precisely than that of corresponding HB stars of clusters.  This opens the possibility of further constraining 
the process competing with atomic diffusion  in causing abundance anomalies.

According to \citet{MorrisonHeSuetal2009} and \citet{KinmanMoBr2009}, inner halo  stars have a mean $[\Fe/\H] = -1.6$ and almost all have $[\Fe/\H] < -0.8$.  
Presumably most field sdB stars have a similar range of original metallicities.    It is then useful to determine to what extent the surface abundances of sdBs should be affected by the various metallicities they may have formed with.  Carrying calculations with different metallicities  allows not only comparisons to globular clusters whose giant branch shows various metallicities but also to analyze potential causes of the abundance range observed in field sdBs.

Among previous work relevant to this paper, one may mention that radiative accelerations of metals and He in sdBs were calculated by \citet{BergeronWeMietal88} and \citet{MichaudBeHeetal89}.  The calculations of \gr(\He) were for stars of 40000\,K or more.
The role of \He{} diffusion for the structure of sdBs has been investigated by \citet{HuNeAeetal2009} and \citet{HuGlThetal2010} but without the important contribution of metal diffusion with radiative accelerations.
    
The  sdBs  correspond to the   highest \teff{}  HB stars observed in globular clusters.  Why did they lose more mass above the He burning core than the cooler HB stars? A number of more or less complicated scenarios have been suggested for them.   Here we assume a simple scenario and determine what anomalies would, in this scenario, be expected on the surface.  We do not claim to exclude more complicated ones but consider useful to determine if surface abundances can be understood in a simple scenario, assuming nature chose that way.

In this paper, the evolution is carried out with models evolved with atomic diffusion from the zero age \MS{} and through the Red Giant Branch, as described in \citet{MichaudRiRi2010}. During the HB, it is continued with a surface mixed zone with enough turbulence to force abundance homogeneity.   The mass of this mixed zone is kept  constant.  Abundance anomalies depend on its extent which is determined by  a comparison to observations (see Sect. \ref{sec:MixedMass}).  In general the calculated anomalies also depend on the original metallicity of the star.  In the case of globular clusters the latter is fixed by the metallicity of the giants of the cluster.  This approach is different from that of \citet{CharpinetFoBretal97} and \citet{FontaineBrChetal2003}, in that these authors assumed the abundances in the exterior regions to be such that gravity and radiative accelerations were at equilibrium.  This led to a parameter free determination of abundances which is a function of \teff{} only.  Their model has considerable success in explaining observations and pulsation properties.  However can the assumed equilibrium always be reached during evolution on the HB?  Is there a sufficient \Fe{} reservoir to fill the region of the \Fe{} opacity bump? Does a mixing process modify it?  Observations suggest a range of abundances for some species in sdBs at a given \teff.

After a very brief description of the calculations (Sect.\,\ref{sec:Calculations}), the internal structure is  analyzed (Sect.\,\ref{sec:internal}), contrasting that of 14000 and 30100\,K stars; the resulting surface abundances are discussed and compared to observations in Sect.\,\ref{sec:Comparison}, insisting on the metallicity dependence of clusters which leads us to continue with field sdB stars (Sect.\,\ref{sec:FieldSdBStars}). After a summary of the main results (Sect.\,\ref{sec:Results}), the potential role of mass loss (Sect.\,\ref{sec:OrMassLoss}) and $\mu$ gradient inversion (Sect.\,\ref{sec:Mu}) are mentioned. While this paper is concerned with the role of stellar evolution for surface abundances, it is concluded by a brief discussion of the need  to test  the models using asterosismology (Sect.\,\ref{sec:Asterosismology}).


\section{Calculations}
\label{sec:Calculations}
Stellar evolution models were calculated  from the zero age \MS{} to the HB  as described in 
\citet{MichaudRiRi2007}.  Using opacity spectra from \citet{IglesiasRo96} all aspects of atomic diffusion transport are treated in detail from first principles. These models are called 
\emph{models with diffusion} in \citet{MichaudRiRi2007}.
   \begin{figure}
   \centering
\includegraphics[width=8cm]{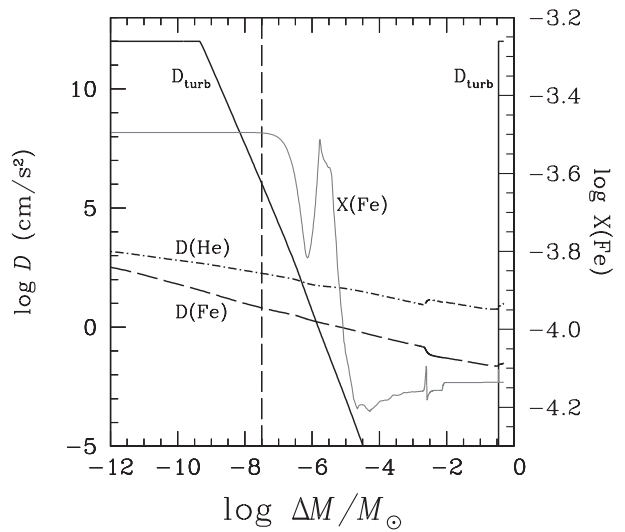}
      \caption{Turbulent diffusion coefficient in a 0.50\,\Msol{} model. The left axis is for the diffusion coefficient and the right axis for the \Fe{} mass fraction.
The coefficient is anchored at a  mass $\Delta M_0 = 10^{-7.5}$\,\Msol{}.   The vertical dashed line is at  $\Delta M_0$. The Fe mass fraction, $X(\Fe)$, is nearly constant from the surface down to $\Delta M = 10^{-7}$\,\Msol{} which corresponds to Fe being mixed down to approximately $3 \times \Delta M_0 $.    }
         \label{fig:Dturb}
   \end{figure}

During the HB, the surface convection zone includes very little mass for $\teff \ge 10000$\,K. To carry out HB evolution, it was found necessary to keep a small surface zone mixed.  A simple turbulent diffusion coefficient (see Fig.\,\ref{fig:Dturb}), similar to those used for AmFm stars of the \MS{}, was used to mix the exterior region during HB evolution.  Turbulence has been assumed large enough to mix completely the regions
between superficial convection zones; such mixing is expected from the results of numerical simulations \citep{KupkaMo2002,FreytagSt2004}.  Below the deepest surface
convection zone, the turbulent diffusion coefficient has been assumed
to obey a simple algebraic dependence on density given, in most calculations
presented in this paper, by
\begin{equation}
     \Dturb= 10^4 D(\He)_0\left(\frac{\rho_0}{\rho}\right)^4 \label{eq:DT}
\end{equation}
where  $D(\He)_0$ is the atomic diffusion 
coefficient of He\footnote{The values of $D(\He)_0$
actually used in this formula were always 
obtained --- for programming convenience ---
from the simple analytical approximation
\hbox{$D(\He)=3.3\times10^{-15}T^{2.5}/[4\rho\ln(1+1.125\times10^{-16}T^3/\rho)]$}
(in cgs units) for He in trace amount in an ionized hydrogen plasma. 
These can differ significantly from the more accurate values used
elsewhere in the code.} at some reference depth. For most calculations in \citet{MichaudRiRi2008},  
the turbulent diffusion 
coefficient was anchored at a given constant temperature, $T_0$.    Then
\begin{equation}
     \rho_0=\rho(T_0)   \label{eq:rho-T-0}
\end{equation}
and equation~(\ref{eq:rho-T-0}) is given by the stellar model.   That density varies during
evolution. 
While a given $\rho_0$ or $T_0$  during the evolution of one star of a given mass on the HB, approximately occurs at a constant value of $\Delta M \equiv \Mstar - M_r$, the latter changes considerably between cool and hot HB  and especially hot sdB stars.  For this paper, most calculations were carried out with 
\begin{equation}
     \rho_0=\rho(\Delta M_0).   \label{eq:Delta-M-0}
\end{equation}
and equation~(\ref{eq:Delta-M-0}) is given by the current stellar model.  In
words, in the calculations reported in this paper, $\rho_0$ of equation~(\ref{eq:DT}) is the density at which
$\Delta M = \Delta M_0$ in the evolutionary model.  For $\Delta M_0 = 10^{-7.5}$\,\Msol{} the concentration of most species is constant for $\Delta M \lta 10^{-7.0}$\,\Msol{} when the assumed turbulent coefficient is $10^4 \times$ the \He{} diffusion coefficient at $\Delta M_0$ and varying as $\rho^{-4}$.  As one increases $\Delta M_0$, one  defines a one parameter family of models.  As may be seen from Fig.\,\ref{fig:Dturb}, turbulence decreases very rapidly inward and using such turbulence is equivalent to mixing a certain mass\footnote{It was found by \citet{RicherMiTu2000} that using 2 or 3 instead
of 4 as the exponent for $\rho$ in equation~(\ref{eq:DT}) only changed the mixed mass corresponding to $\Delta M_0$.
It was further found by \citet{TalonRiMi2006} that distinguishing between different turbulence models required an accuracy of 0.03\,dex which is beyond the accuracy of current abundance determinations (see their Sect.\,5).}.
The exact mass over which concentration is kept uniform by turbulence depends on the driving terms in the atomic diffusion equation and so varies slightly with atomic species.  As may be seen from Fig.\,\ref{fig:Dturb}, it is $3 \times \Delta M_0$ for \Fe{}, but it is $10\times \Delta M_0$ for \He{} (not shown).  For $ \log (\Delta M_0/\Msol) = -7.5$ these models are labeled dM--7.5D10K-4.

   \begin{figure*}
   \centering
\includegraphics[width=9cm]{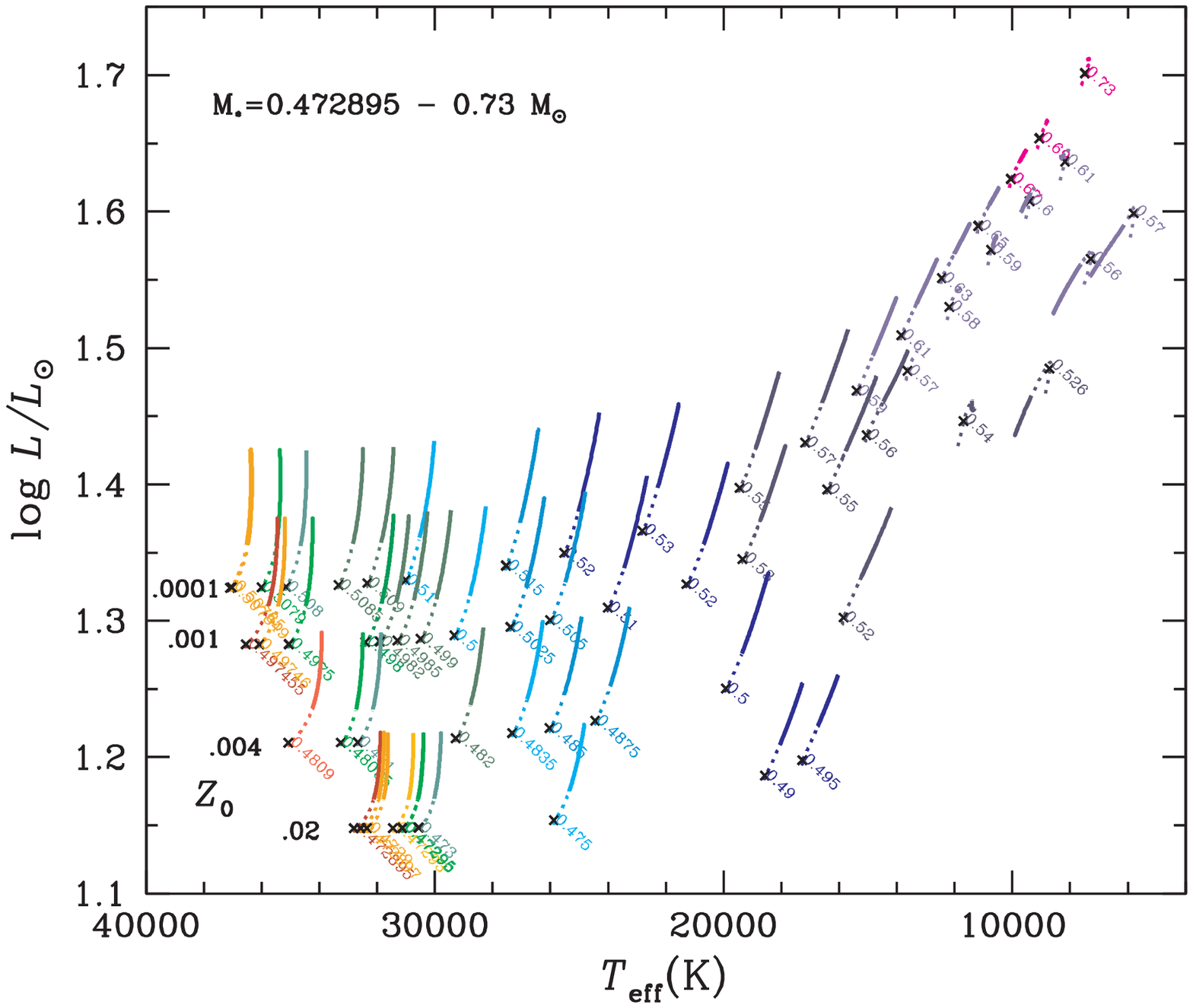}
\includegraphics[width=9cm]{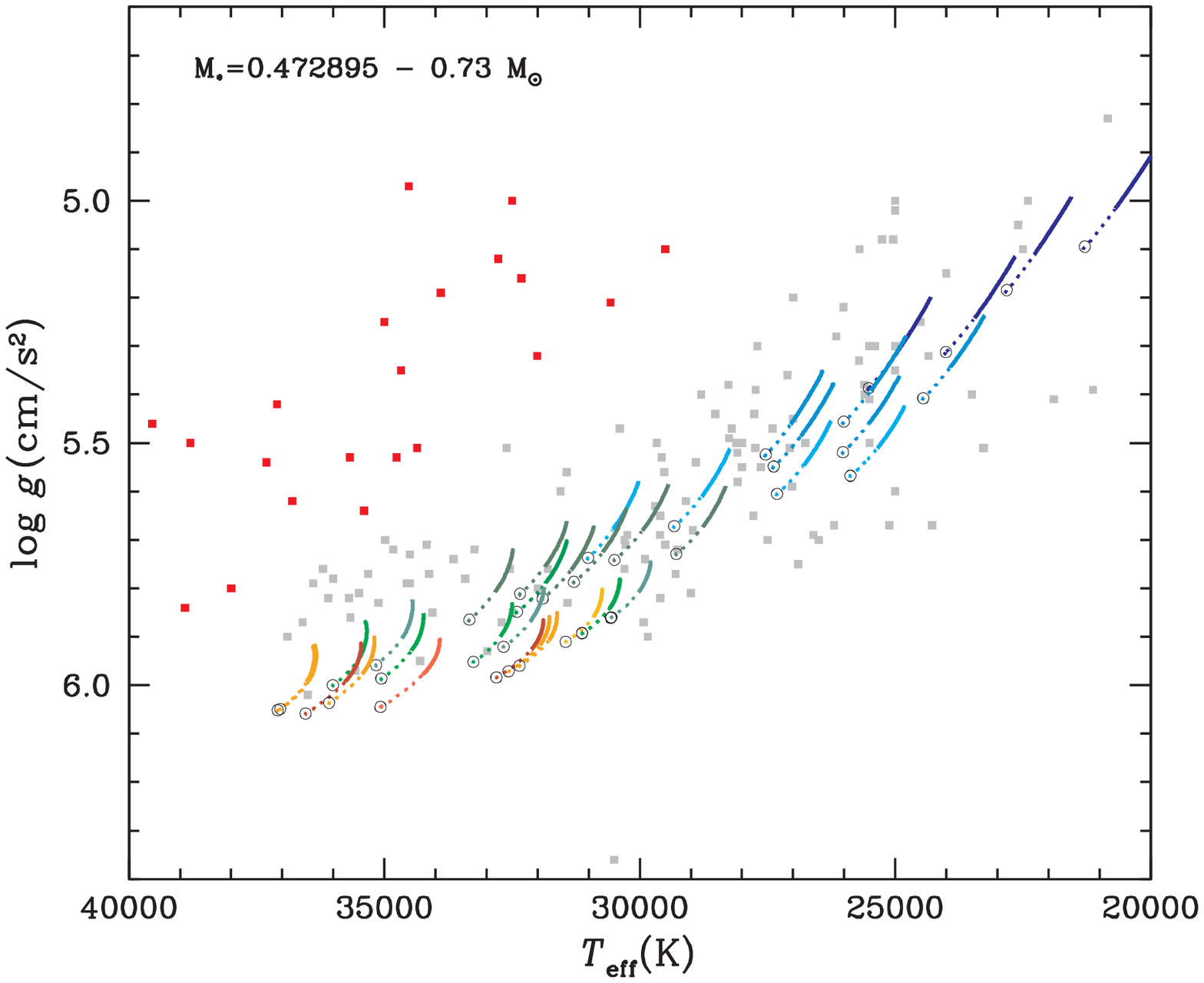}
      \caption{HR diagram (left)  for all calculated models with a mixed mass $\sim 10^{-7} \msol$.    Each segment represents the evolution of one model whose mass is color coded using a simple formula. The metallicity ($Z_0$) of each series of models is indicated at the left of the sequence. The dotted part represents the first 10\,Myr of HB evolution while the solid line represents the rest of the calculated HB evolution, ususally an additional  27\,Myr.  In the right panel,  g vs \teff{}  diagram of a subset of the calculations (those with $\teff > 20000$\,K) corresponding to the observations of  \citealt{GeierHeEdetal2010}.   This panel contains data (gray and red dots as defined in the text) from their Fig.\,2.   }
         \label{fig:DMfixed}
   \end{figure*}
      \begin{figure}
   \centering
\includegraphics[width=8cm]{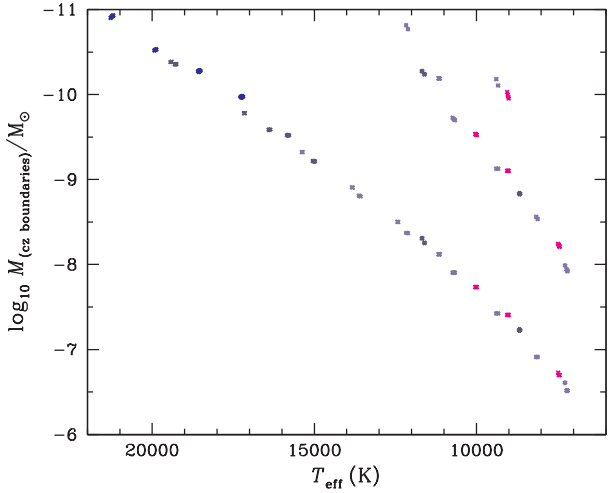}
      \caption{Mass within and above the He II, He I and H convection zones as a function of \teff.   The H convection zone is separate from the He I convection zone in only a few models. For stars hotter than 22000\,K, the temperature at the bottom of the convection zone is less than a factor of 2 larger than \teff{} and is probably not determined precisely by interior models; it is not shown.  The color code is the same as in Fig.\,\ref{fig:DMfixed}.}
         \label{fig:MBCZ}
   \end{figure}

      \begin{figure*}
   \centering
\includegraphics[width=9cm]{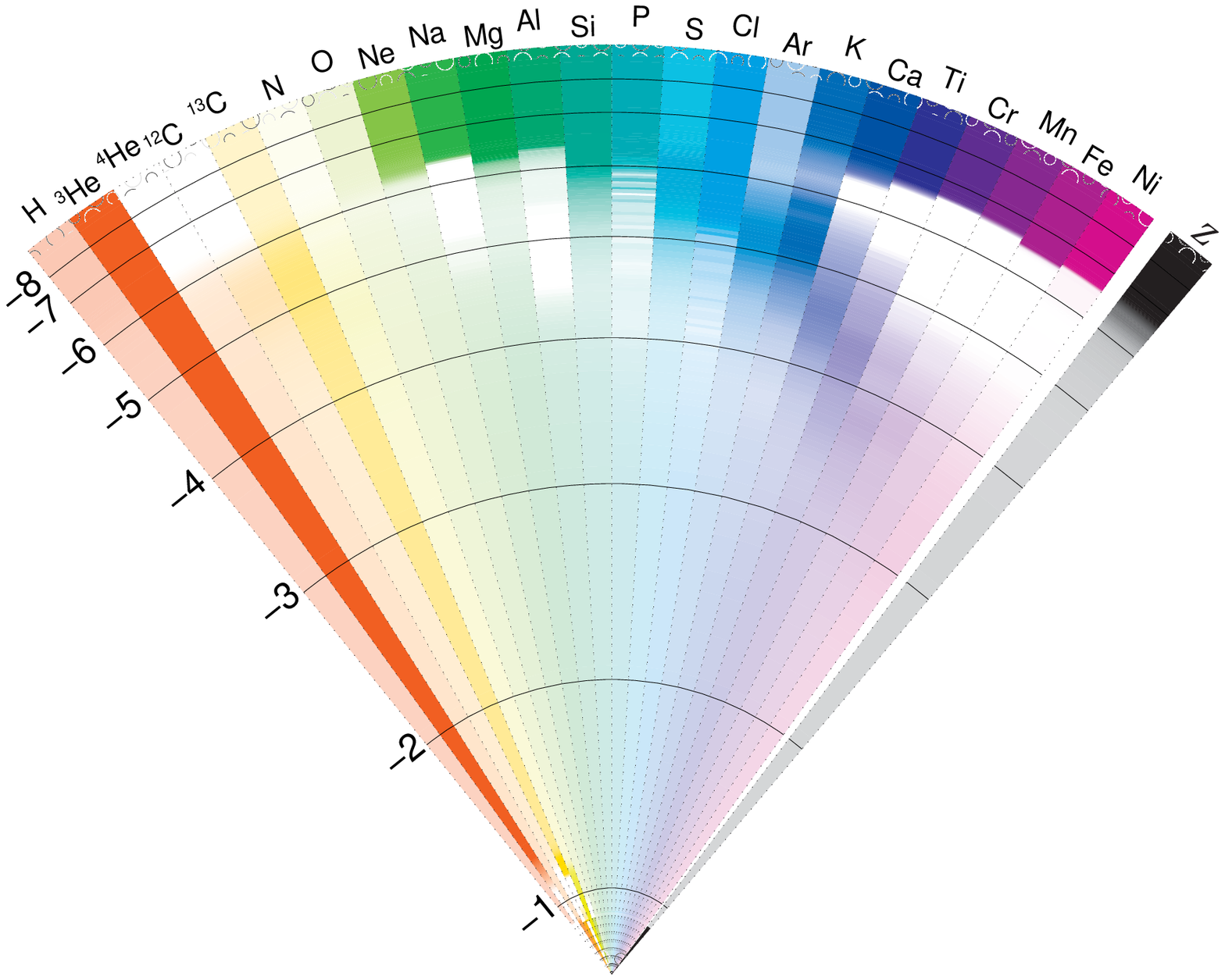}
\includegraphics[width=9cm]{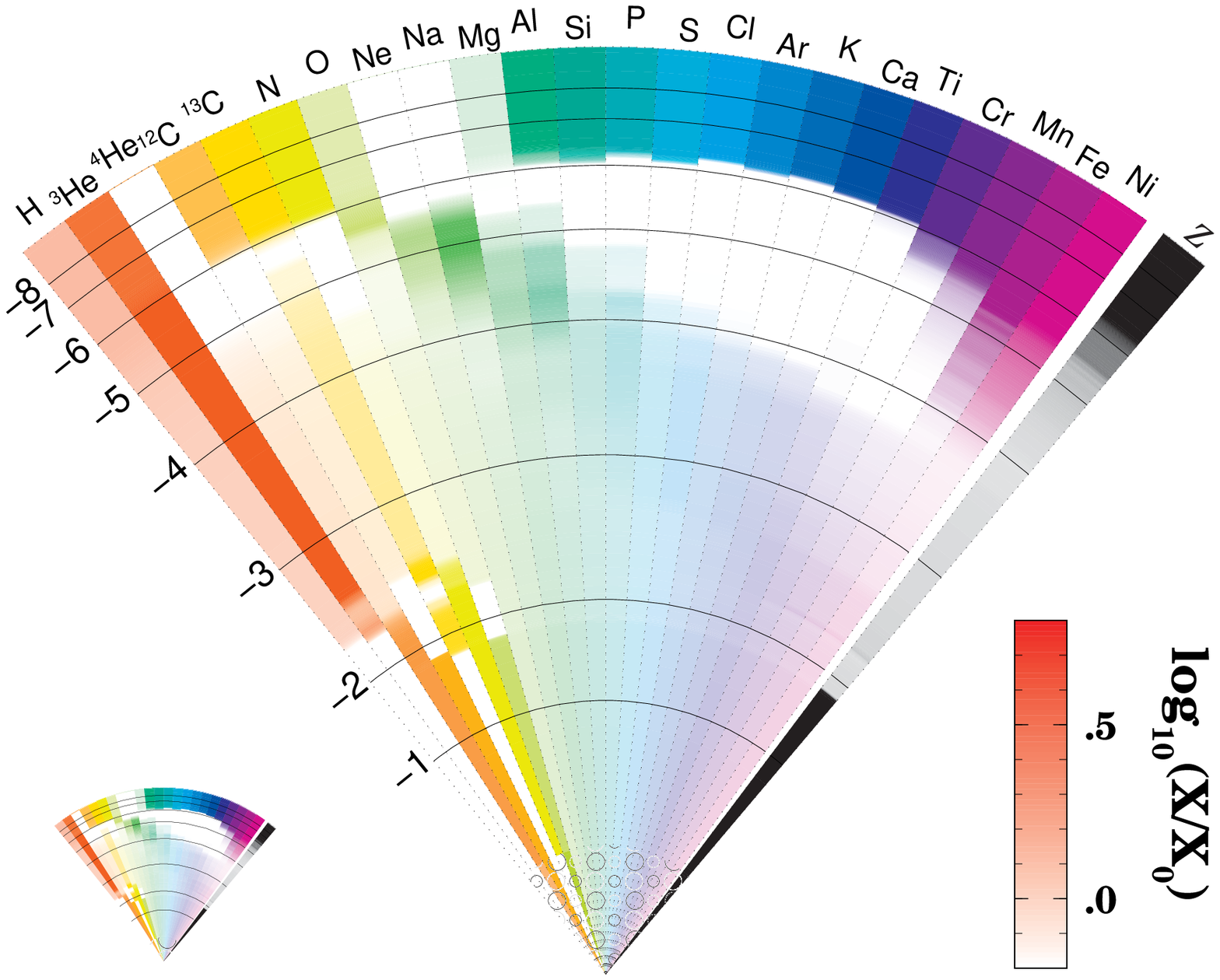}
      \caption{Color intensity coded concentrations in two HB stars of the same metallicity after 25\,Myr on the HB. Left panel with a \teff{} of  14\,000\,K (0.59\,\msol{}) and right panel of 30\,100\,K (0.51\,\msol).  The radial coordinate is the radius and its scale  is linear but the logarithmic value of the mass coordinate above a number of points, \DM, is shown on the left of the 
	horizontal black line.  The concentration scale is given in the right insert.  
	  Small circles near the top of the left panel
	mark the extent of the surface convection zone while similar circles near the center of both models mark the central convection zone.   The small inset in between the two panels shows the high \teff{} star, that is the right panel, on the radius scale of the low \teff{} star, that is the left panel.    For $-7<\DM < -4$ the concentration is quite different for many species.  It is surprisingly so for C and O for $\DM>-2$.  See the text. A black and white version of this figure may be found in the On line Fig.\,\ref{fig:EventailsBW}.}
         \label{fig:Eventails}
   \end{figure*} 
      \begin{figure*}
   \centering
\includegraphics[width=9cm]{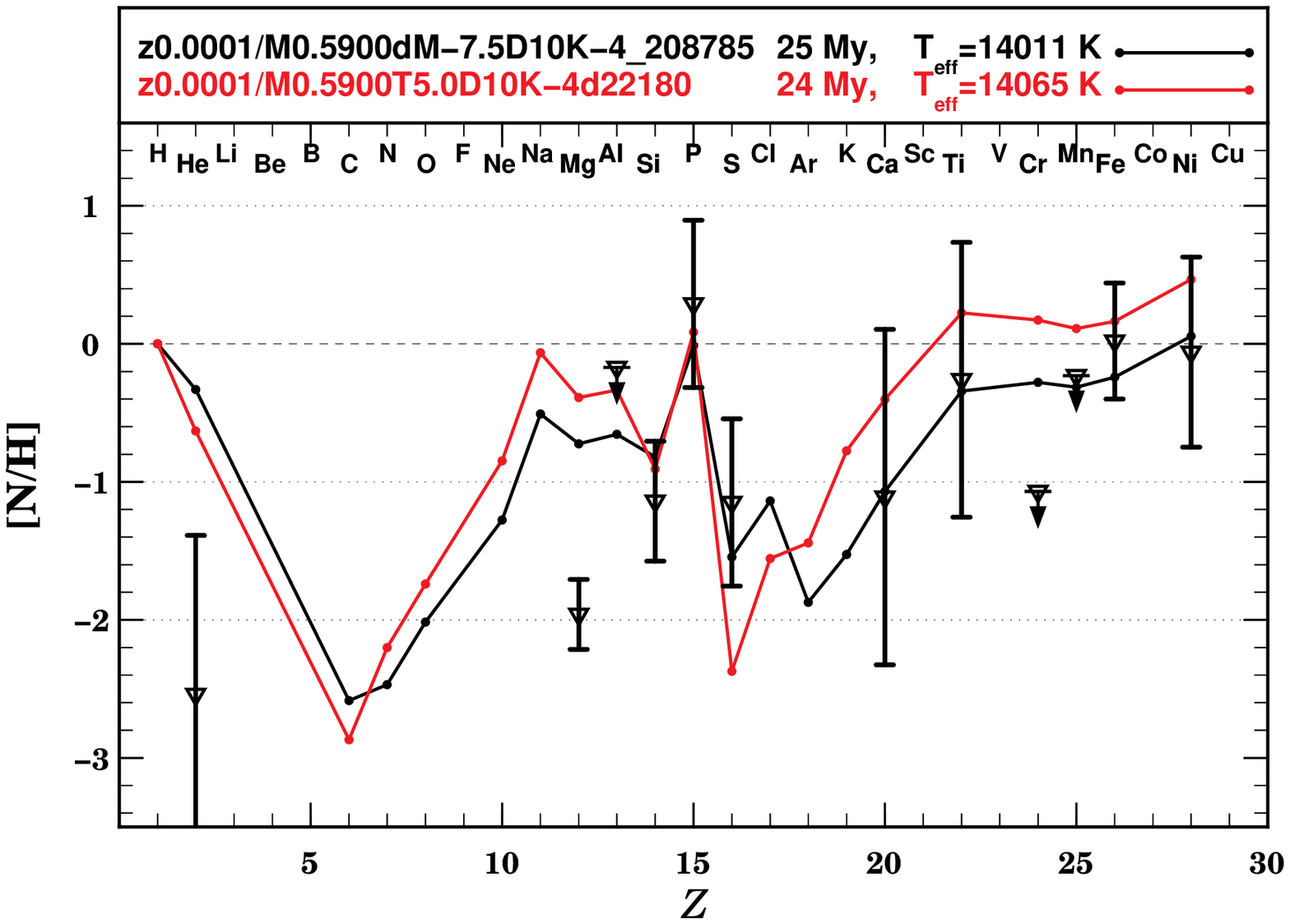}
\includegraphics[width=9cm]{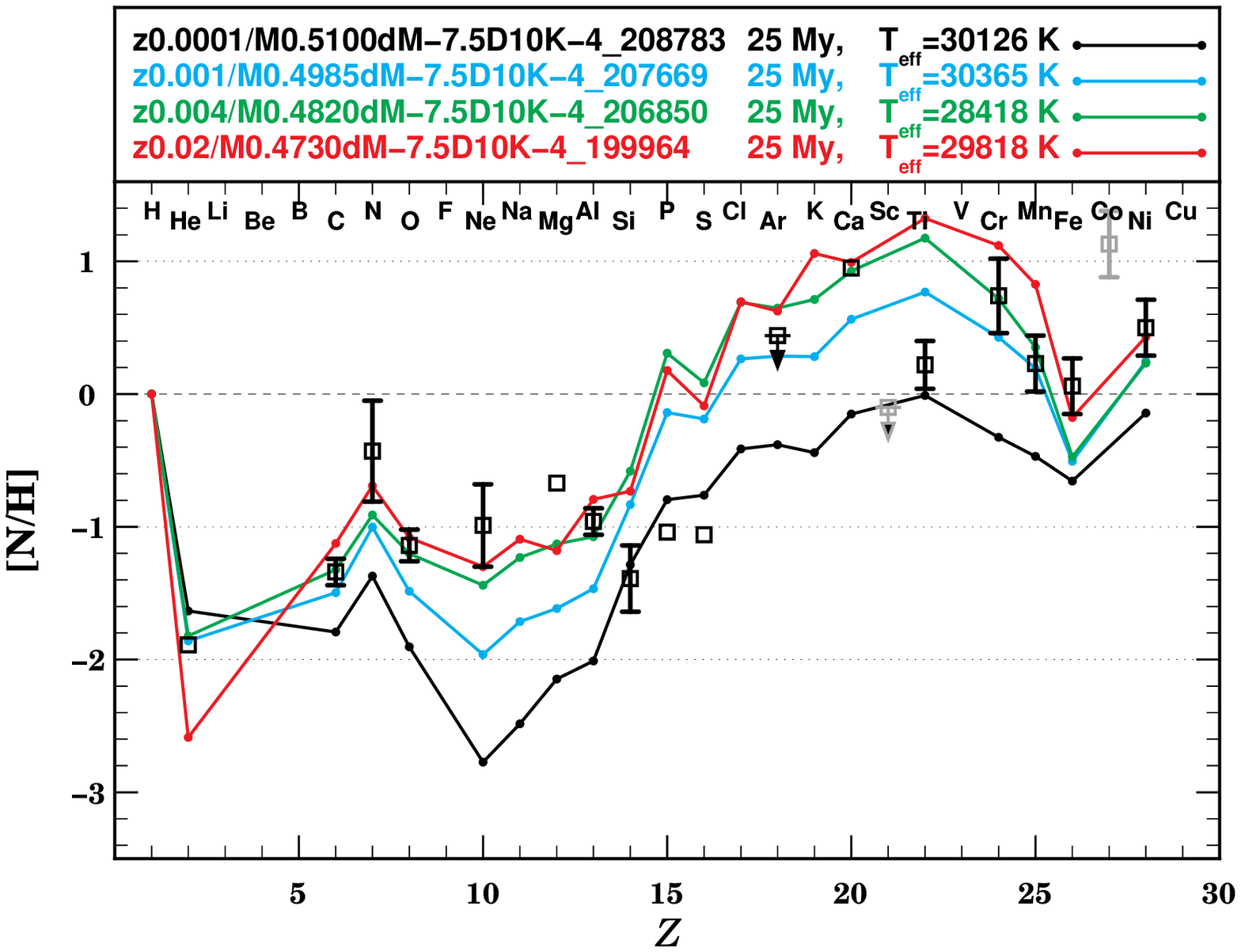}
      \caption{Comparison of surface abundances expected in 14\,000 (black line, left panel) and 30\,100\,K (black line, right panel) HB stars with $ Z_0 = 0.0001$ whose interior is shown in Fig.\,\ref{fig:Eventails}. On the left panel, in order to compare with the turbulence model used by \citet{MichaudRiRi2008} (see the text for details), the red  curve is taken from the left panel of their Fig.\,11   and the data is for B203 from \citet{Behr2003}.  On the right panel the other curves correspond to stars of  original metallicities of $Z_0 = 0.001$, 0.004   and 0.02 and the data  for Feige\,48 is from \citet{OtooleHe2006}. Grey is used for data points  of atomic species which are not included in our calculations because they are not in OPAL opacities.}
         \label{fig:2surfaces}
   \end{figure*}
     
     \begin{figure*}
   \centering
\includegraphics[width=11cm]{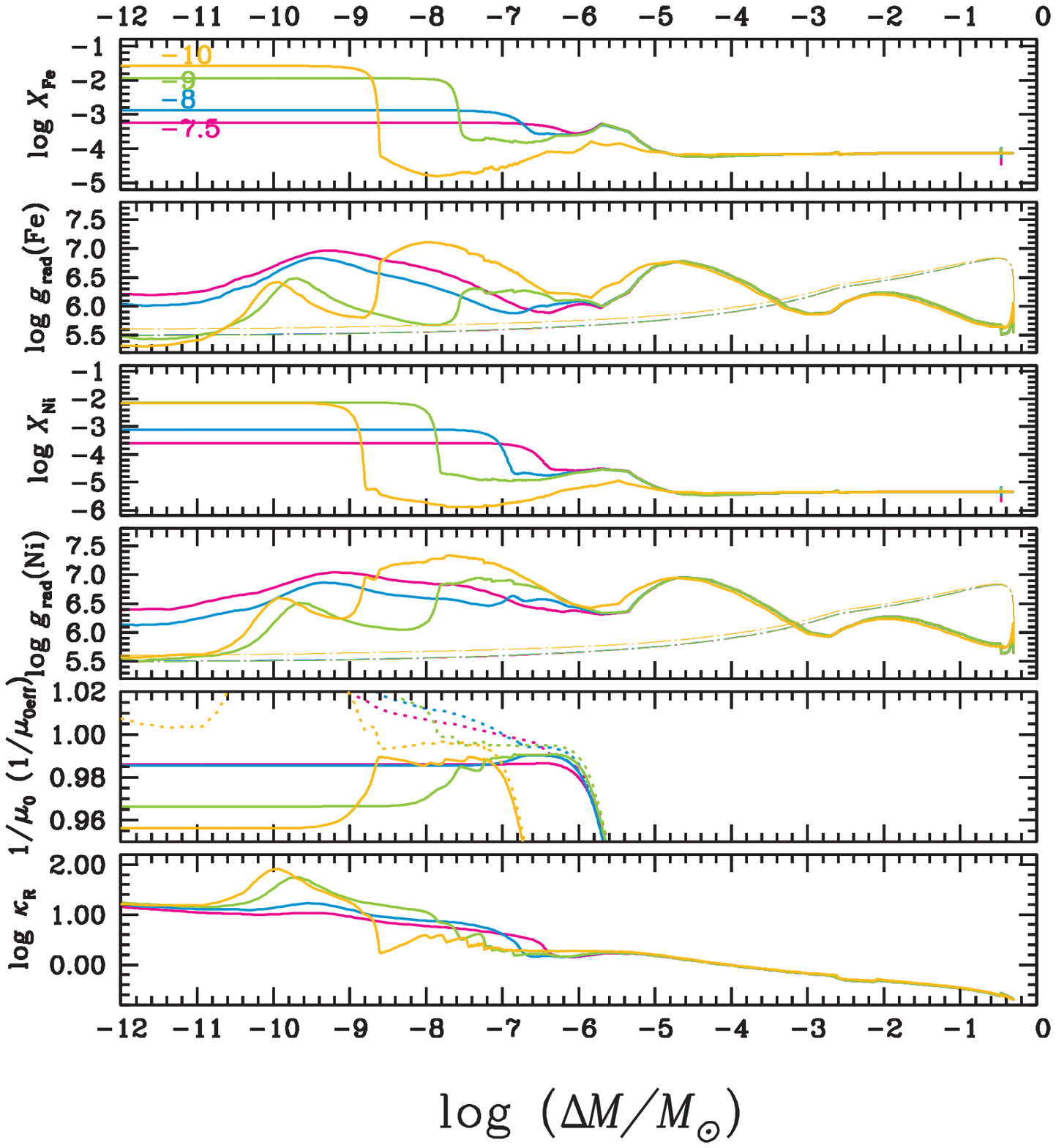}
      \caption{Internal distribution of \Fe{} and \Ni{} together with their relation to \gr(\Fe) and \gr(\Ni) in a 0.5\,\msol{} model  with $Z_0 = 0.001$ for four   turbulence strengths, each anchored at a fixed $\Delta M_0$ (see Eq.\,[\ref{eq:DT}] and [\ref{eq:Delta-M-0}]), identified on the upper panel. The solutions labeled $-7.5$, $-8$ and $-9$ are after 32 Myr on the HB while the solution labeled $-10$ is after 7\,Myr. The lower panel gives  $\kappa_{\mbox{\scriptsize R}}$ as a function of \DMsol{}. A local  maximum of \gr(\Fe) and of $\kappa_{\mbox{\scriptsize R}}$ occurs for \DMsol{} between -10 and -9 and corresponds to $\log T\sim 5.5$. 
    When $\Delta M_0 \lta 5\times  10^{-8} \msol$ the iron abundance becomes rapidly much larger than solar in contradiction with observations of sdB stars.  
The function $1/\mu_0$  is shown (solid lines) for each model  in the panel above that for $\kappa_{\mbox{\scriptsize R}}$.  In the same panel, the function  $1/\mu_{0\mathrm{eff}}$ is shown by dotted lines.  It is defined in Sect.\,\ref{sec:Mu}.}
         \label{fig:Fe_gr}
   \end{figure*}  
   
\section{Abundance anomalies and internal concentration variations}
\label{sec:internal}
Calculations were carried out for metallicities of $Z_0 =$\,0.02, 0.004, 0.001 and 0.0001.   Corresponding tracks in the HR diagrams and in the $\log g$--\teff{} plane
are shown in Fig. \ref{fig:DMfixed}.  The time evolution of $L$, \teff, of surface abundances of \He, \Ca, \Cr{} and \Fe, and of other variables of interest may be found as an on line Appendix (App.\,\ref{sec:Appendix}) for all models.

 On the right panel of Fig.\,\ref{fig:DMfixed} the data from  Fig.\,2 of \citet{GeierHeEdetal2010} is compared with the $g$ vs \teff{} covered by our  models\footnote{\label{foot:color}We use  color codes that depend on  the envelope mass, that is on $\log_{10}(M_* - M_{\mbox{\scriptsize{He core}}})/M_{\odot}$.  This expression is used to define an index referring to 16  predefined colors. The same code is used for all figures involving  stars with $\teff > 20000$\,K. Another code is used when only stars with $\teff < 20000$\,K are present.
  }. 
These cover the range of most stars in that paper for $\teff < 36\,000$\,K.  There are some lower gravity stars that have gravities more than a factor of 2 smaller than those of our models; they are identified using a different color (red).  According to Fig.\,1 of \citet{CharpinetFoBretal2000}, they should corresppond to the end of the HB phase which is not treated in our models.  When we compare the results to  Figs.\,3 and 4 of \citet{GeierHeEdetal2010}, and Fig.\,1 of \citet{GeierHeNa2008} we will usually identify them separately. Evolution was stopped  when the He mass fraction dropped below $\sim 60$\,\%  in the star center (after around 32\,Myr on the HB) because the algorithm used to calculate diffusion velocities requires the presence of either He or H.  For technical reasons eliminating this limitation requires the rewriting of major sections of the code and this is outside the scope of this paper. For a few models the calculations were stopped after less than 32 Myr (see App.\,\ref{sec:Appendix}) because of convergence problems.

The mass in the surface convection zone is usually one of the most important stellar model properties for atomic diffusion since it is thoroughly mixed and separation generally occurs below.  The bottom of the He II, He I and H convection zones are shown in Fig.\,\ref{fig:MBCZ}. The bottom of the He II convection zone  does not depend significantly on metallicity in so far as the points define a single line (for instance, at low \teff, the magenta stars are on the same line as the blue stars). Just as occurs on the \MS{} of Pop II stars (see Fig\,1 of \citealt{RichardMiRi2005}), the mass in and above the surface convection zone  depends nearly only on  \teff{} which effectively defines the depth of the surface convection zone.  Calculated models cooler than 22000\,K are shown with metallicities ranging from $Z_0=0.0001$ to 0.02.  For stars hotter than 22000\,K, the temperature at the bottom of the convection zone is less than a factor of 2 larger than \teff{} and such outer structure is probably not determined precisely by interior models. It was decided not to show the mass above that \teff.  The mass mixed by turbulence, $\sim 10^{-7}$\msol{} in this paper (see Sect.\,\ref{sec:MixedMass}), is always larger than the mass in the surface convection zones.  The latter then does not play a major role.

\subsection{A 30100\,K sdB vs a 14000\,K HB star}
\label{sec:sdb vs HB}
   On Fig.\,\ref{fig:Eventails}, are compared the internal concentrations in a 30\,100\,K HB (0.51\Msol; right panel) to those in a 14\,000\,K HB star (0.59\Msol; left panel).  Both started with the same metallicity ($Z_0 = 0.0001$) and have spent the same time on the HB, 25\,Myr.  Given that they both started with $Z_0=0.0001$, the two  have similar \Fe{} peak surface overabundances (a factor of $\sim$\,200 overabundance from the original abundance of the star) but very different He, N, Ne, Na, Mg, Al, and Si surface abundances (compare the black curves in the right and left panels of Fig.\,\ref{fig:2surfaces}). However the lower temperature model has a radius 5.4 times larger than that of the hotter model. The hotter model is shown on the radius scale of the cooler one in the small central inset.  The central region (defined as the region where $^{12}$C has been produced by \He{} burning) of the two models is quite similar.  By  comparing the central region of the inset to the central region of the left panel, one notices that the various concentrations are about the same at approximately the same radius and inner mass (that is as a function of $m_r$, the mass within radius $r$) in both models.  	However while abundance variations caused by nuclear reactions extend over the inner 50\,\% of the radius in the hotter model, they extend only over the inner 10\,\% of the radius in the cooler model. Furthermore the central region corresponds to rather different stellar mass fractions of the two models: it is the inner 99\,\% by mass of the 0.51\Msol{} star but it corresponds to the inner 85\,\% by mass of the 0.59\,\Msol{} model.  The difference originates from the different peeling applied to the red giant in order to get to the different \teff{} of the two models:   14\,\% more mass  has been removed from the red giant to get to the 30\,100\,K model than to the cooler model.  In fact approximately the same inner mass is affected by nuclear reactions in both models and the structure of the central core is similar in both.  However the envelope has quite a different structure,  since it extends much further in the higher mass model.  Abundance variations caused by diffusion are seen to extend over the outer 40\,\% by radius for many species in both models   so that, as a function of the fractional radius as used in Fig.\,\ref{fig:Eventails}, the outer region ($ \DM{} < -5$) has many similarities in the two. However between  $ \DM{} = -5$ and the central region, the concentration variations are  quite different.  Furthermore even when the fractional radius is about the same in the two models, the thickness in physical distances differs by the scale factor of 5.4.

   On Fig.\,\ref{fig:2surfaces} the black curves represent the surface abundances of the two stars of Fig.\,\ref{fig:Eventails}:  \emph{left panel},  0.59\,\Msol{} model; \emph{right panel},  0.51\,\Msol{} model.  The right panel also contains results for models of larger metallicities than used in Fig.\,\ref{fig:Eventails}; the original metallicities range up to $Z_0= 0.02$.
 The surprise is the similarity of the surface Fe abundance: it is nearly solar in both the left and right panels and for all curves of the right panel; this corresponds to a factor of 1 to 200 overabundance from the original abundance of the star depending on the original metallicity.  The Fe abundance seems to be determined by  saturation of \gr(\Fe) where  the separation occurs.     One may relate them to the \gr(\Fe) at the bottom of the mixed zone for the turbulence model used.  The black curves in the left and right panels are quite similar from Ca to Ni but differ strongly from C to K.
   This figure will be further discussed in Sect.\,\ref{sec:Comparison}.

\subsection{Determining the mixed mass}
\label{sec:MixedMass}
On Fig. \ref{fig:Fe_gr}, are shown the internal distributions of Fe and Ni in four models of the same mass and original metallicity but with different turbulence.  The original Fe mass fraction at the beginning of HB evolution was $\sim 10^{-4.1}$ throughout the star with only small variations left over by red giant evolution (see \citealt{MichaudRiRi2010}).  The \gr(\Fe) has  clearly defined peaks at  around $\DMsol = -7.5$ and at around $\DMsol = -5$ (these may also be seen in Fig.\,3 of \citealt{MichaudRiRi2008}).  In between these peaks \gr(\Fe) remains slightly larger than gravity allowing a limited \Fe{} flux to be pushed toward the surface and creating a small underabundance over the interval $-5 < \DMsol <-4$, the iron from that mass interval being pushed toward the surface.  Between $\DMsol = -5$ and slightly below the bottom of the mixed zone, the \Fe{} abundance is determined by the requirement of flux conservation.  Immediately below the mixed zone, an \Fe{} gradient develops to satisfy the diffusion equation as  \gr{}(\Fe), the Fe abundance in the mixed zone and the Fe flux coming from below evolve.
The largest turbulence shown is labeled by $\log \Delta M_0/\Msol = -7.5$; it is used for most calculations and leads to  about solar surface Fe abundance.
The surface Fe abundance is mainly determined by the value of \gr(\Fe) immediately below the mass mixed by turbulence.  Since at $\log T \sim 5.4$, \gr(\Fe)  is larger than $g$ even at solar \Fe{} concentrations, mixing only from the surface to $\DMsol = -8.6$ (or $\log T \sim 5.4$) leads to larger $X(\Fe{})$  than observed. One needs to mix further inwards, to $\DMsol \sim -7$ for surface Fe abundances to be about solar.  This in practice is what fixes the mixed zone for the models presented in this paper. The mass in the mixed zone was so fixed by the Fe abundance in sdBs.  It was then used for all sdB and HB calculations with \teff{} from 7000 to 35000\,K.  A mixed mass a factor of 10 smaller would clearly lead to unacceptably large Fe abundances, while a factor of three larger is marginally excluded. 

In the third panel, is shown $1/\mu_0$, where $\mu_0$ is the reduced mass per nucleus, that is excluding electrons. When $\log (\Delta M_0/\msol) < -7.5 $, it has a local maximum interior to  the \Fe{} accumulation.   It is caused by the accumulation of metals pushed upwards by \gr{}.  The local maximum is very small in the case $\log (\Delta M_0/\msol) =-7.5 $.  There still is an accumulation of metals in this case above $\DMsol = -6.0 $ but its effect in increasing $\mu_0$ is largely canceled, for $\log (\Delta M_0/\msol) =-7.5$, by the decrease of the \He{} concentration. 
It was verified (not shown) that in the case of turbulence with $\log (\Delta M_0/\msol) = -7.5 $, there was no $\mu_0$ inversion related to metal overabundances for metallicities of $Z_0 = 0.0001$, 0.001 and 0.02 and, in each case, in both a star of 30000\,K and of 15000\,K.  The case shown in Fig.\ref{fig:Fe_gr} is the one with the largest effect among all those verified.

   \begin{figure*}
   \centering
\includegraphics[width=0.45\textwidth]{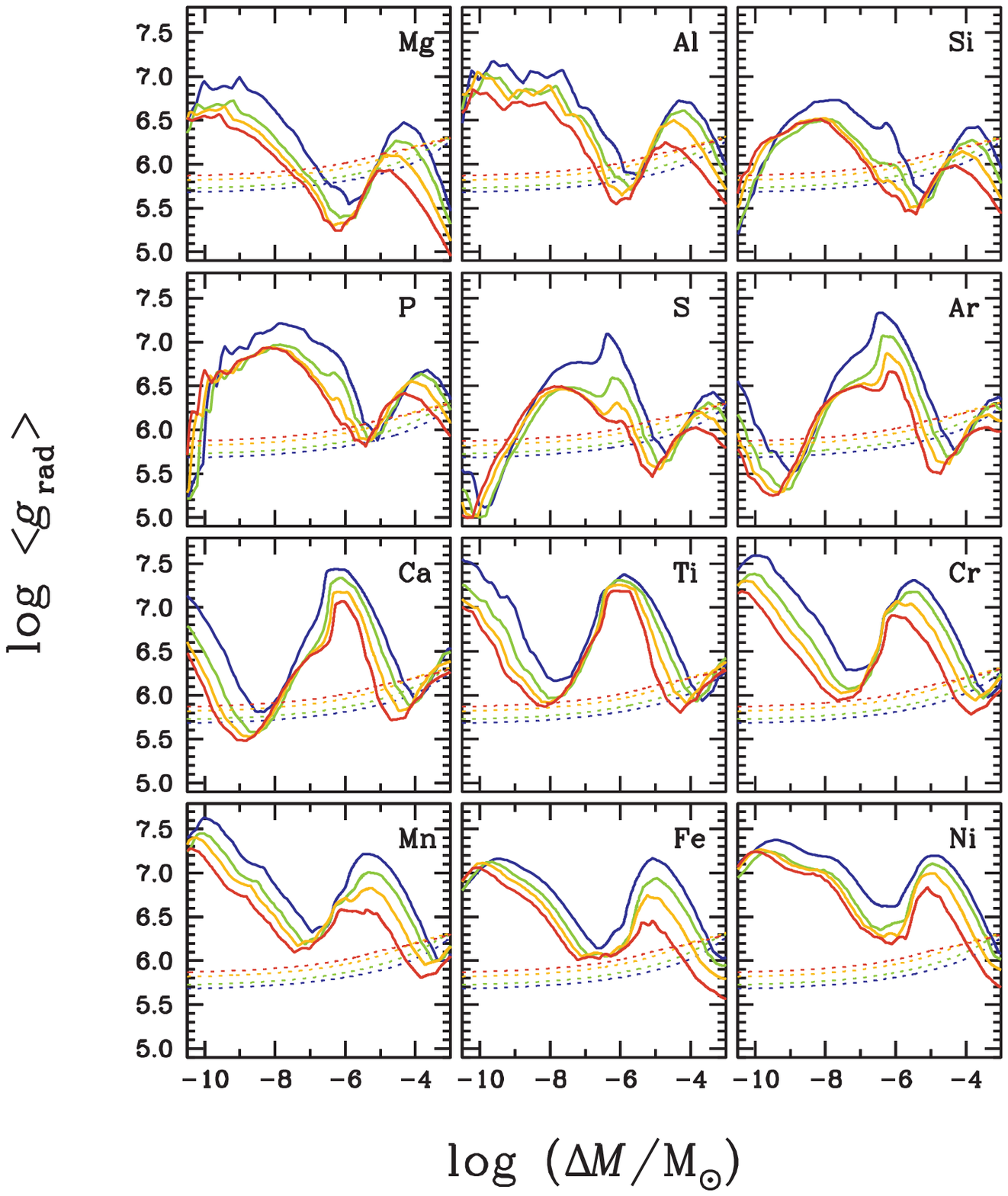}
\hspace{\fill}
\includegraphics[width=0.45\textwidth]{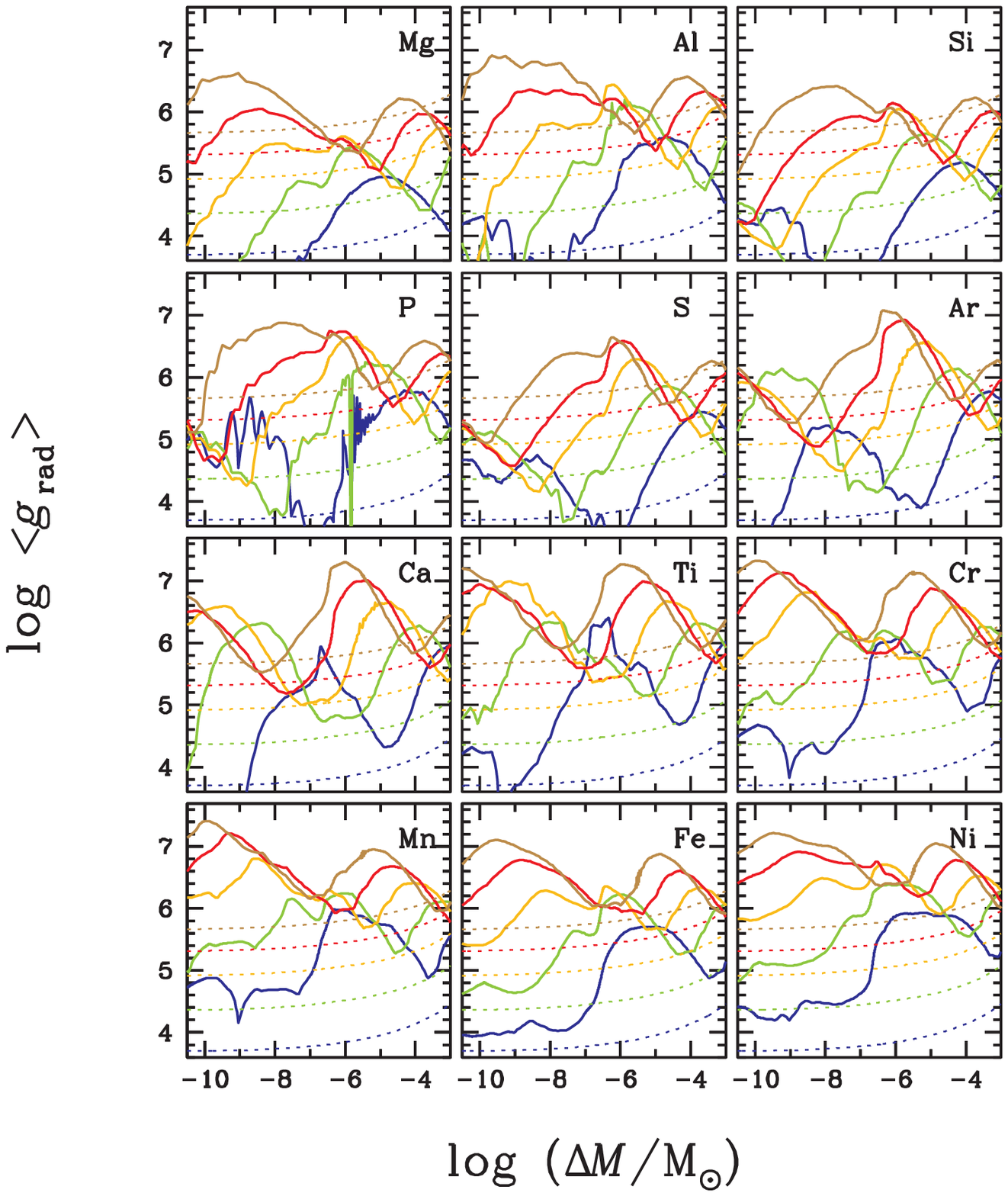}
      \caption{ Radiative accelerations for a subset of the included species in  models with turbulence anchored at a fixed $\Delta M_0 = 10^{-7.5}$\,\Msol{}, after approximately 25\,Myr on the HB.  Dotted lines represent gravity.  Left panel: in  models with $\teff \sim 31000$\,K and metallicities of $Z_0= 0.0001$, 0.001, 0.004 and  0.02, (masses from 0.51 to 0.47 \msol) from top to bottom.
 Right panel: with $Z_0 = 0.001$ and various masses corresponding, from bottom to top, to \teff{}s of 10\,700, 15\,000, 20\,000, 25\,000 and 30\,400\,K.     
           }
         \label{fig:gr}
   \end{figure*}
    \begin{figure*}
   \centering
\includegraphics[width=0.45\textwidth]{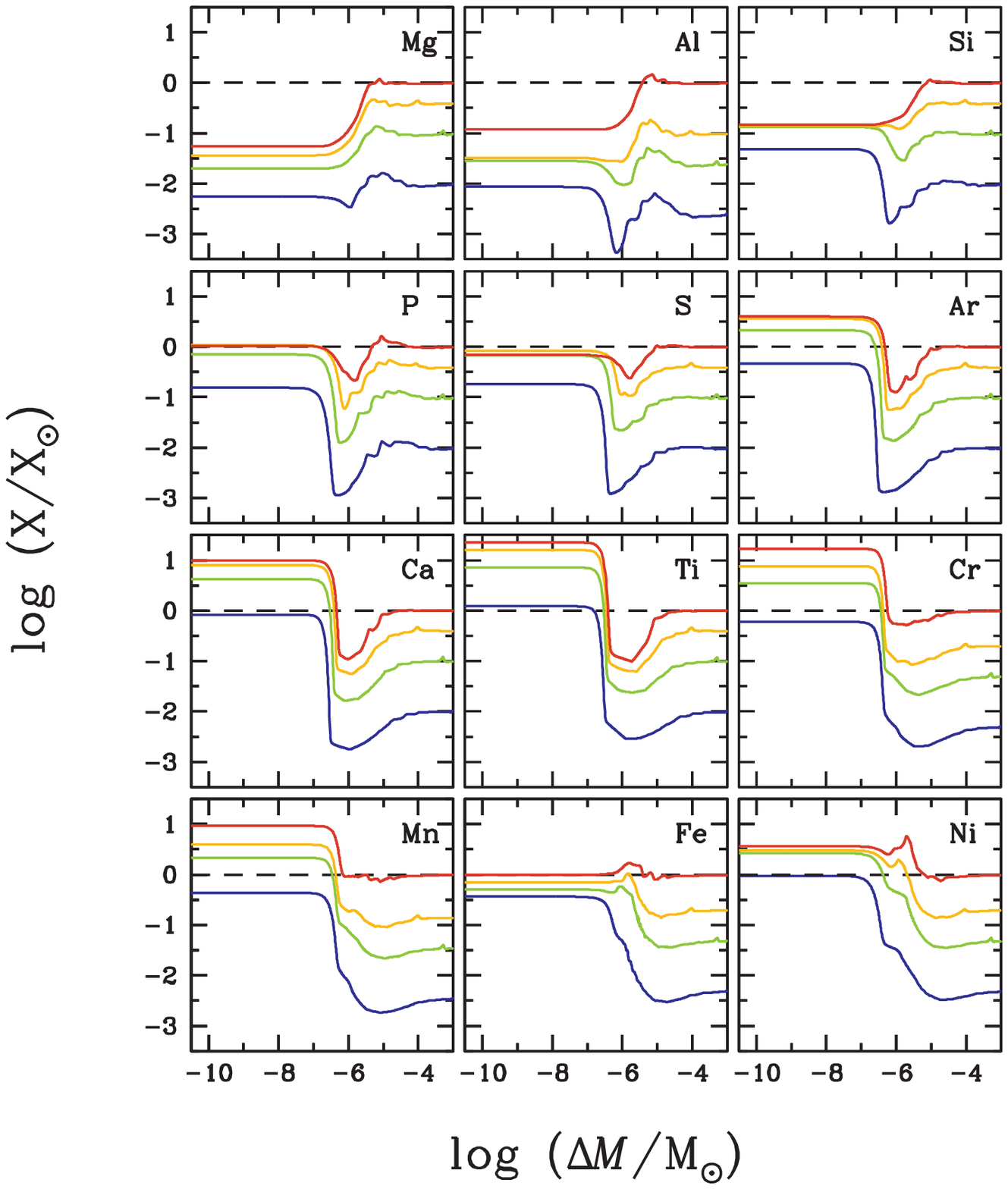}
\hspace{\fill}
\includegraphics[width=0.45\textwidth]{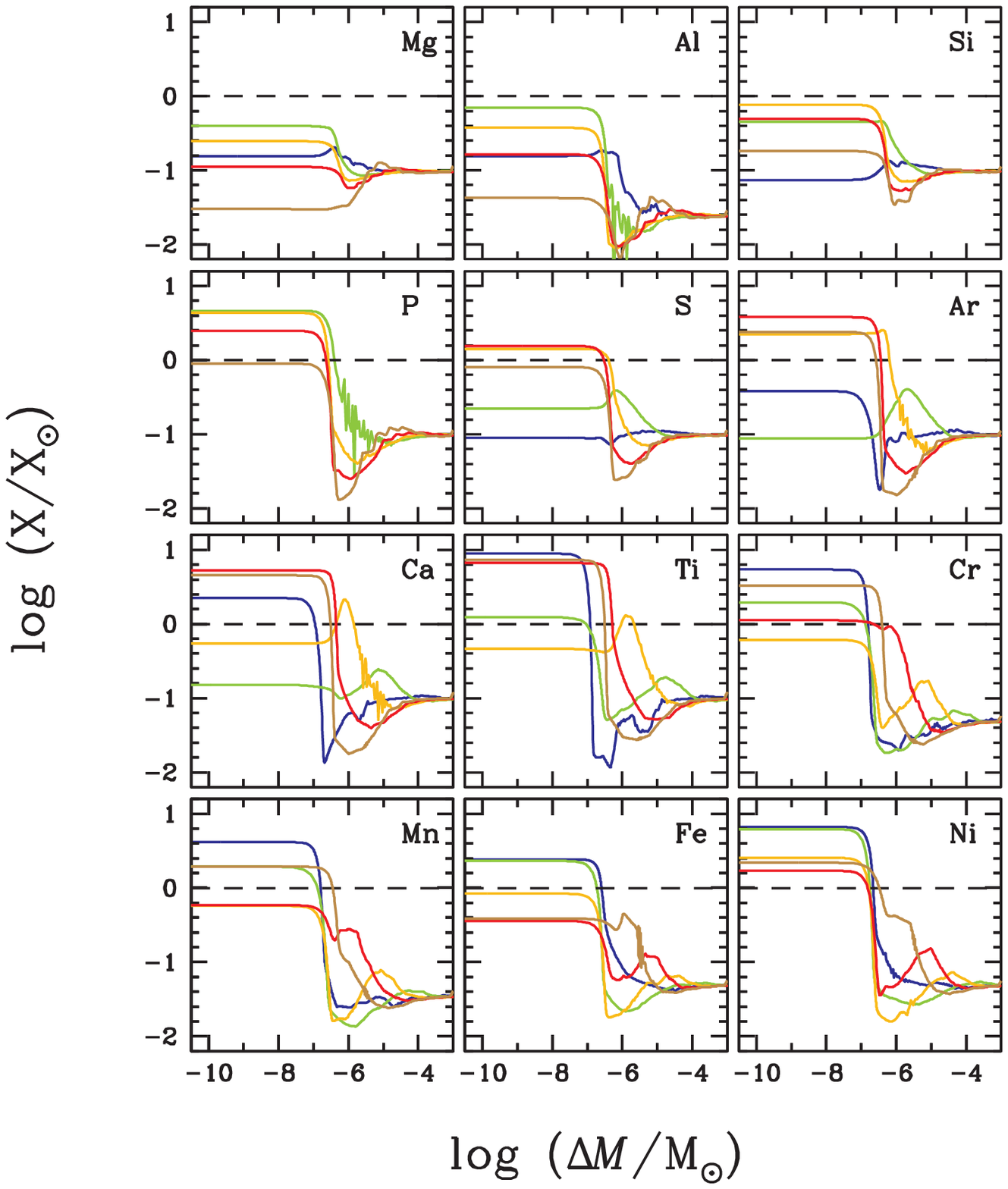}
      \caption{Abundance profiles corresponding to the \gr{}s of Fig.\,\ref{fig:gr}.  The abundance at the right of each small panel still approximately equals the starting abundance while that at the left is the surface abundance after 23\,Myr of HB evolution. 
           }
         \label{fig:interiorX}
   \end{figure*}
 
\subsection{Radiative accelerations and interior mass fractions}
\label{sec:gRad}
Radiative accelerations  are shown at $\sim 25$\,Myr after ZAHB in Fig.\,\ref{fig:gr}; in the left panel for various metallicities   while in the right panel, they are shown at various \teff{}s for a metallicity of $Z_0= 0.001$.  The \gr{}s clearly vary more with \teff{} at a given metallicity than with metallicity at a given \teff{}. The change with metallicity is mainly due to saturation of the lines which causes a general reduction of \gr{} as metallicity is increased; also as metallicity is increased, $T$ increases at a given $\log (\Delta M_0/\msol)${} and a given ionization state shifts towards the surface.  As \teff{} increases however, the radiative flux increases with $\teff^4$ and furthermore, gravity increases by orders of magnitude, strongly affecting the ionization equilibrium at a given  $\log (\Delta M_0/\msol)$ within the star.  As seen on the figure, a given \gr{} maximum  migrates toward the surface as \teff{} increases.
 
In the right panel of Fig.\,\ref{fig:gr}, one notices large scatter  for \gr(P) in the $\teff = 10700$\,K model. We have verified by an analysis of the spectra used to calculate \gr(P) that very few lines contribute to it in the $\teff = 10700$\,K model.  At $\log (\Delta M_0/\msol) = -6$, where the \gr(P) curve is most irregular, a single line contributes at least 30\,\% of the value and 5 lines, the following 30\,\%.  Given that these are calculated in opacity sampling, this leads to the observed fluctuations.  For more details, see the discussion of \gr(Li) in \citet{RicherMi2005} and in Sect.\,3 of \citet{VickMiRietal2010}.  Because of the uncertainties in \gr(P), the corresponding curve for $X(P)$ is not shown in Fig.\,\ref{fig:interiorX}.  The \gr{}s only matter where turbulence does not force homogenization so only for $\log (\Delta M_0/\msol) > -7$.
No large glitches due to improper sampling occurs for any of the other cases shown  for $\log (\Delta M_0/\msol) > -7$.  Abundance variations as seen in Fig. \,\ref{fig:interiorX} do lead however to variations in \gr{}s.  In the same model, for instance, the \Ca{} abundance minimum below the mixed zone is partly responsible for the \gr(\Ca) peak at the same place.  The large overabundance of \Ni{} in the mixed region of the $\teff = 10700$ and 15\,000\,K  models are largely responsible for the rapid variation of \gr(\Ni) there.

In the left panel of Fig.\,\ref{fig:interiorX}, one sees that the surface abundances at 25\,Myr increase as the starting abundances increase.  However, saturation plays a role for all species shown: the range in surface abundances is smaller than the range of starting abundances.  Saturation is most evident for Si and Fe, whose surface abundances differ by at most a factor of 3 while starting abundances covered a range of 200, still visible in the deeper layers.  In the right panel of Fig.\,\ref{fig:interiorX}, one sees a much less regular behavior.  The starting abundances were the same for all models as still evident from the deeper layers, at $\log (\Delta M_0/\msol) = -4$.  The surface abundance of Si is smallest in the coolest model, but the surface abundance of Ni is largest in the coolest model.  Iron peak elements have a similar behavior with \teff.  Lower mass nuclei vary much more differently (compare Al and Ca for instance) as could be expected from the behavior of \gr{}s on the right panel of Fig.\,\ref{fig:gr}.   The shape of the surface abundance curve as a function of $Z$, the nuclear charge number, is then expected to vary with \teff{}.
     \begin{figure*}
   \centering
\includegraphics[width=9cm]{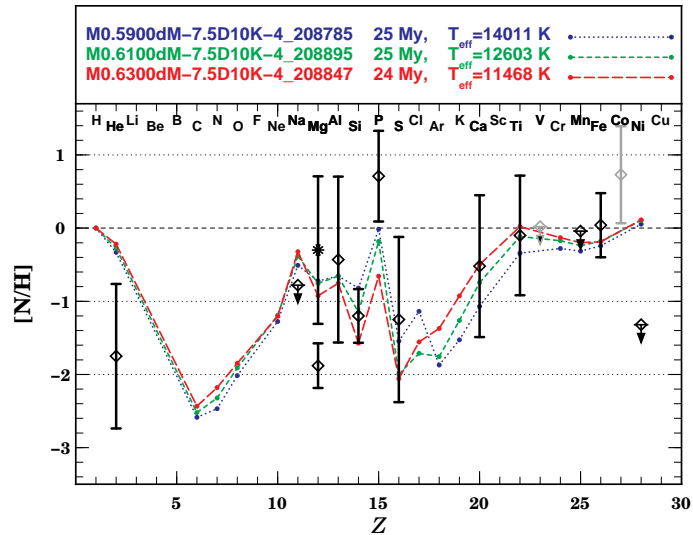}
      \caption{Surface abundances expected for models  with turbulence anchored at  $\Delta M/\msol =10^{-7.5}$, $Z_0\sim  0.0001$ and three slightly different masses, or equivalenty \teff{}.    
      Observations for  star B315 of M\,15 with $\teff\sim 13\,000$\,K are from \citealt{Behr2003}.  It is the same observations as used in the right panel of Fig.\,11 of  \citet{MichaudRiRi2008}.  The turbulence models are however different.  The agreement is equivalent.      }
         \label{fig:M15}
   \end{figure*}
   \begin{figure*}
   \centering
\includegraphics[width=9cm]{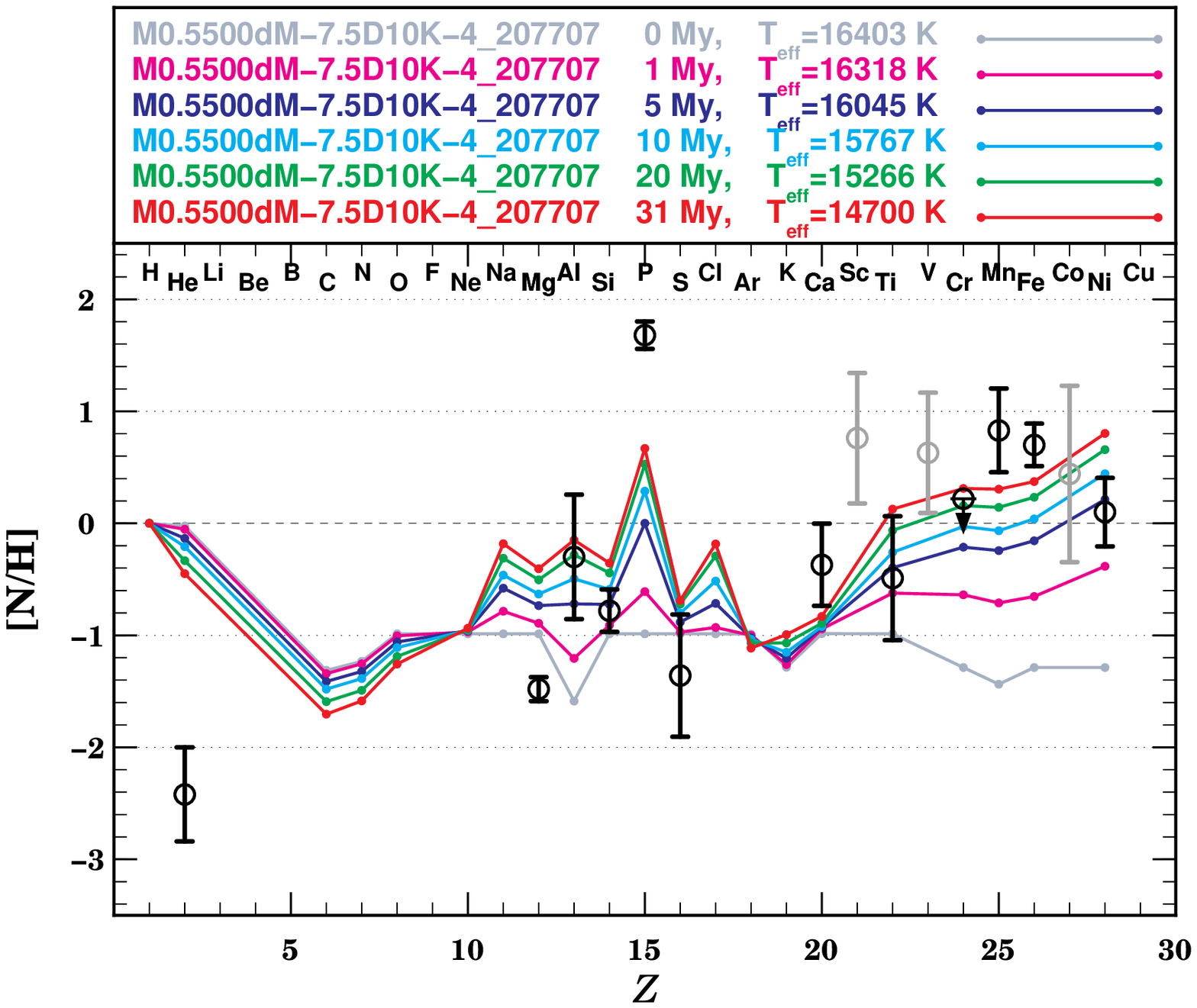}
\includegraphics[width=9cm]{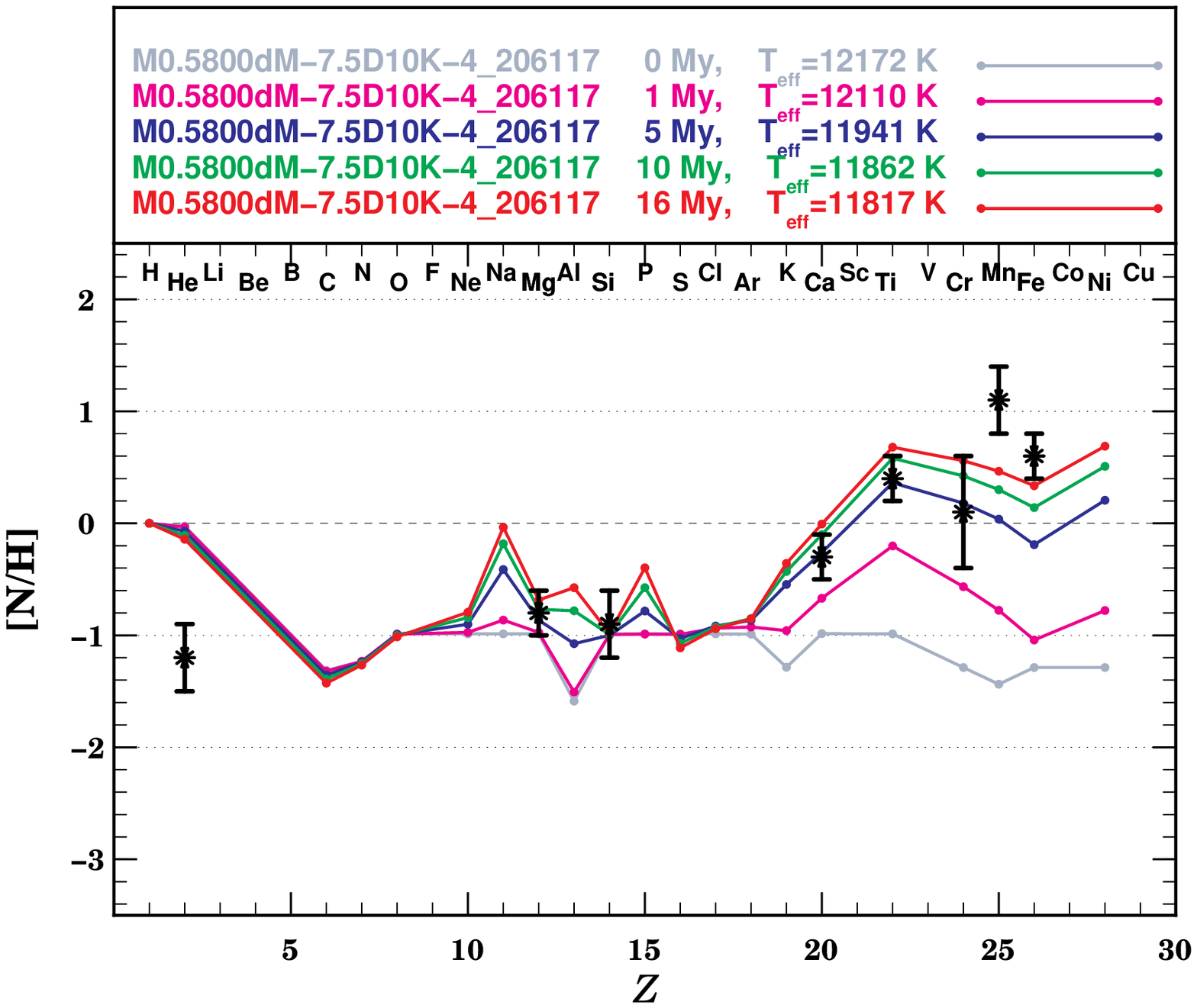}
      \caption{Comparison of abundance patterns with models of metallicity [Fe/H]$=-1.3$ and with turbulence anchored at  $\Delta M/\msol =10^{-7.5}$. \emph{Left panel} Surface abundances expected at six HB ages, from 0.0 to 31\,Myr, for a 0.55 \msol{} model   compared to    
      observations for WF4--3085, a star of M\,13 with $\teff\sim 14\,000$\,K,  from \citealt{Behr2003}.   
      \emph{Right panel}  Surface abundances at five HB ages, from 0.0 to 16 Myr, for a 0.58\,\msol{} model compared to observations of the $\teff =$ 12000\,K star 469 of NGC1904 from \citet{FabbianReGretal2005}.    }
         \label{fig:M13}
   \end{figure*}
     \begin{figure*}
   \centering
\includegraphics[width=8cm]{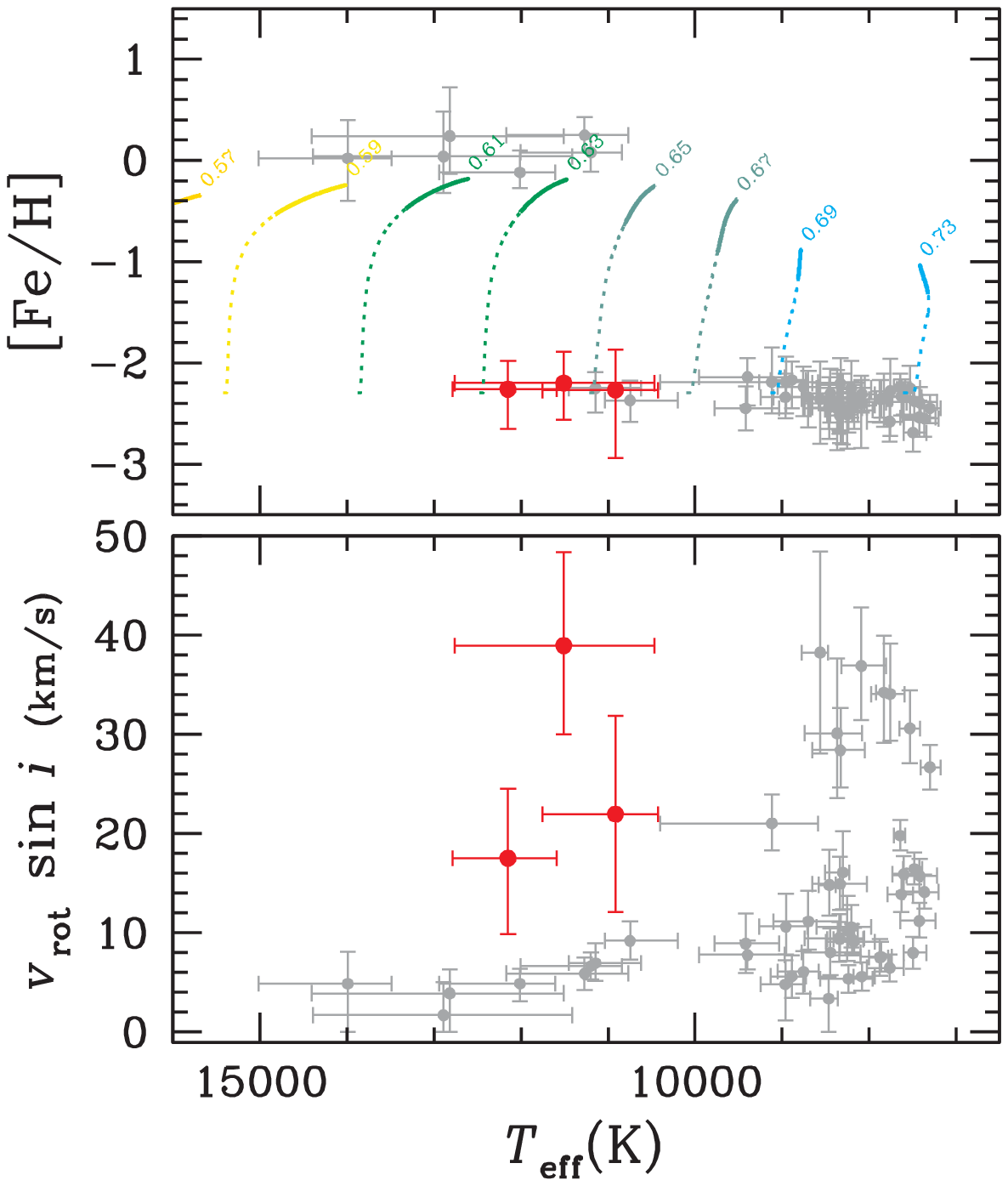}
\includegraphics[width=8cm]{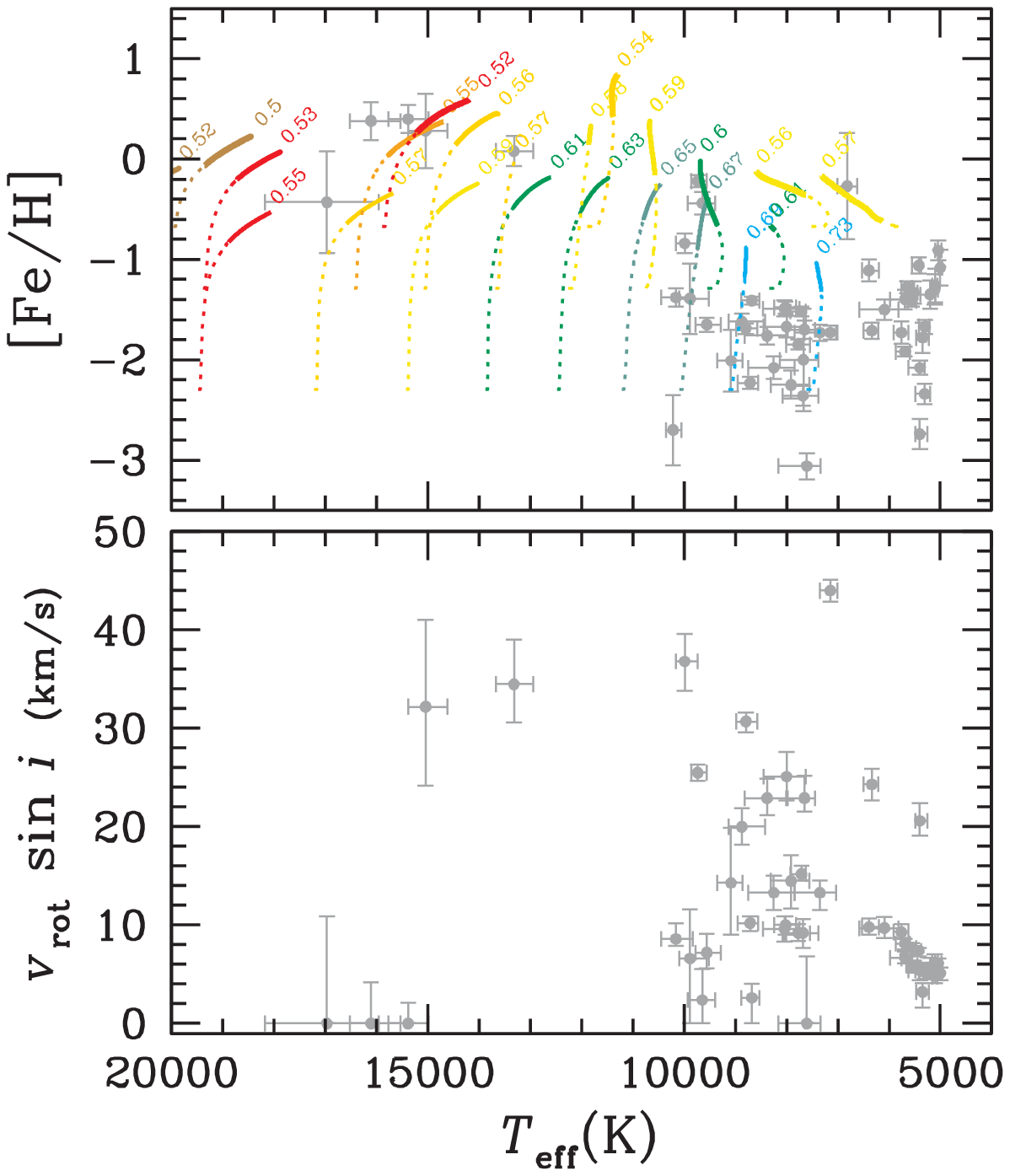}
      \caption{  
       \emph{Left panel} Surface concentration of Fe as a function of \teff{} in clusters (M15, M68 and M92) from \citealt{Behr2003} with original Fe concentration $\sim$-2.3 . Models are shown for  this metallicity. Each  color coded segment represents the evolution of the surface Fe concentration. The line is solid for the time interval from 10 to 30 Myr after ZAHB but dotted from 0 to 10 Myr.   Bottom part: rotation velocities; three stars have their rotation velocities and [Fe/H] in red to facilitate their identification.
      \emph{Right panel} Field stars also from  \citealt{Behr2003}.  The distribution of metallicity for the cooler field stars gives an indication of the original metallicity distribution for the hotter field stars as well as for sdBs.  }
         \label{fig:Fe0.0001}
   \end{figure*} 
   
\section{Comparison to Observations}
\label{sec:Comparison}
\subsection{Metallicity, \teff{}, and age dependence}
\label{sec:Dependence}
Before comparing to observed abundance patterns of individual stars, it is useful to evaluate the dependence of calculated abundances  on original metallicity, \teff{} and age.

If one observes a star of 30000\,K, how sensitive are its surface abundances to the original stellar metallicity?  In  the right panel of Fig.\,\ref{fig:2surfaces} the surface abundances at 25 Myr after ZAHB are shown in four stars of original  metallicity ranging from $Z_0 = 0.0001$ to 0.02.  They were chosen to have \teff{} as close as possible to 30000\,K since \teff{} is usually the better determined quantity. This imposed different masses which range from 0.47 to 0.51 \msol.  While the original abundances vary by a factor of 200, the final ones vary by at most a factor of 30 and for \Fe{} and Ni by a factor of about 3.  The largest differences are for species around Ca and  Ne.  Generally a star that starts with a larger metallicity ends up having larger abundances of metals, but there are a few exceptions such as P and S in the $Z_0 = 0.004$ and 0.02 models. This must be caused by a competition for photons with, for instance, Fe (see Fig.\,4 of \citealt{RicherMiRoetal98}).  Helium is most underabundant in the higher $Z_0$ stars.

Comparing the right and left panels of Fig.\,\ref{fig:2surfaces} shows that a factor of 2 difference in \teff{} for stars originally with $Z_0= 0.0001$ does not change considerably the expected \Fe{} peak abundances but leads to large differences for atomic elements with masses equal to or smaller than that of  Ca.   While a peak occurs for N in the higher \teff{} model, it occurs for P in the lower \teff{} one.  The differences can reach 1--2 orders of magnitude.  However for smaller \teff{} differences, the abundance variations tend to be relatively small (see Fig.\,\ref{fig:M15}).  Phosphorus is most sensitive, its abundance variation reaching a factor of 5 between 11500 and 14000\,K (however see also Sect.\,\ref{sec:gRad}).  But for Fe and Ni an uncertainty of 2500\,K hardly affects the expected surface abundances even though they are more than a factor of 100 overabundant.  By comparing the left and right panels of Fig.\,\ref{fig:M13} one notes that a relatively small \teff{} change ($\sim 3000$\,K) can lead, in a cluster with $Z_0= 0.001$, to significant abundance variations for species between Si and Ti.  For heavier iron peak elements, the variations are much smaller.

Examples of  age dependence are shown on Fig.\,\ref{fig:M13}.  The abundance variations are  rapid for the first 5 Myr but much smaller between 10 and 30 Myr for all species whose abundance increases.  The variations are much more regular for species with $A< 10$ which become underabundant.  In a population of stars of various unknown HB ages, one consequently  expects most stars to have at least the overabundance attained after the first 5 Myr. See also Figs.\,\ref{fig:modelsZ0001} to \ref{fig:modelsZ02}.

\subsection{Low metallicity clusters}
\label{sec:Low}
In \citet{MichaudRiRi2008}  one turbulence model anchored at a fixed $T$ was shown to reproduce  abundance anomalies   observed on the HB of M\,15 from 11000 to 15000\,K.  A single turbulence model was used for all HB stars of that cluster.  For this paper, we tried using the same model for HB stars of higher metallicities and for sdBs of higher \teff{} but  it was found not to reproduce all observations.  We then searched for a model anchored at a fixed $\Delta M/\msol$,  so at a fixed  fraction of \Msol.  
It is \emph{a priori} not clear how the different turbulence models (given by the difference between Eq. \ref{eq:rho-T-0} and \ref{eq:Delta-M-0} during evolution) should affect surface abundances.
It was first verified that one obtains  with turbulence anchored at  $\Delta M/\msol =10^{-7.5}$,  as good an agreement between observations and calculations as shown on Fig.\,9 to 12 of \citet{MichaudRiRi2008} by comparing the detailed abundances for 2 stars of M\,15 and by comparing the \Fe{} abundance for all stars as a function of \teff.

Results obtained with both turbulence models, are compared on
the left panel of Fig.\,\ref{fig:2surfaces} to  observations of the $\teff \sim 14000$\,K star B203, from \citet{Behr2003}.  For all calculated species, the surface abundances calculated with both turbulences are within a factor of 3 of each other except for S where the agreement with observations is better with the model of this paper.  Note that for iron peak elements the abundances are  more than 100$\times$ the original abundances.

In Fig.\,\ref{fig:M15}, observations \citep{Behr2003} of star B315 of M\,15  ($\teff\sim 13\,000$\,K) are compared to calculations with turbulence anchored at $\Delta M/\msol =10^{-7.5}$ for stars of slightly different masses or \teff{}. Two observed values are shown for Mg, the upper one for Mg\,I and the lower one for Mg\,II.  The agreement may be compared to that obtained in the right panel of Fig.\,11 of \citet{MichaudRiRi2008}.    At least from their effect on surface composition, the two turbulence models are seen to be equivalent for B315.  

A third comparison with previous results may be made by comparing Fig.\,9 of \citet{MichaudRiRi2008} to  the left panel of Fig. \ref{fig:Fe0.0001}, where observed \Fe{} abundances in the  clusters M\,15, M\,68 and M\,92 are compared with our calculations over the \teff{} interval from 7000 to 15\,000\,K.  Lines represent the Fe surface abundance evolution during the first 30 Myr of HB evolution. They are color coded as a function of the mass of the model, the first 10 Myr, dotted, the following 20 Myr, solid.  Given the slope of the solid lines, the Fe abundance may continue to increase by a factor of 1.5--2 during the following 60 Myr.  The calculated abundances are then in agreement with observed abundances from 11\,000 to 15,000\,K.  The stars with $\teff < 11000$\,K are influenced by rotation as shown by \citet{QuievyChMietal2009}.
 The comparison to observations is extended to more clusters of low metallicity than considered by \citet{MichaudRiRi2008}.  Few stars hotter than 11000\,K are added but there are three rapidly rotating ones; they follow the pattern explained by \citet{QuievyChMietal2009} and are shown in red. When compared to the results of Fig.\,9 of \citet{MichaudRiRi2008} those of Fig.\,\ref{fig:Fe0.0001} vary less as \teff{} increases from 11000 to 16000\,K.  There is a hint that this agrees better with observations, though the error bars are too large to allow a firm claim.

The model used here for all calculations reproduces observations  as well as that used in \citet{MichaudRiRi2008}.  
 As will  be verified in Sect.\,\ref{sec:FieldSdBStars}, fitting observed anomalies of stars over a large \teff{} interval requires mixing to involve a given mass, approximately the same independent of metallicity and \teff{}. 
   \begin{figure*}
   \centering
\includegraphics[width=8cm]{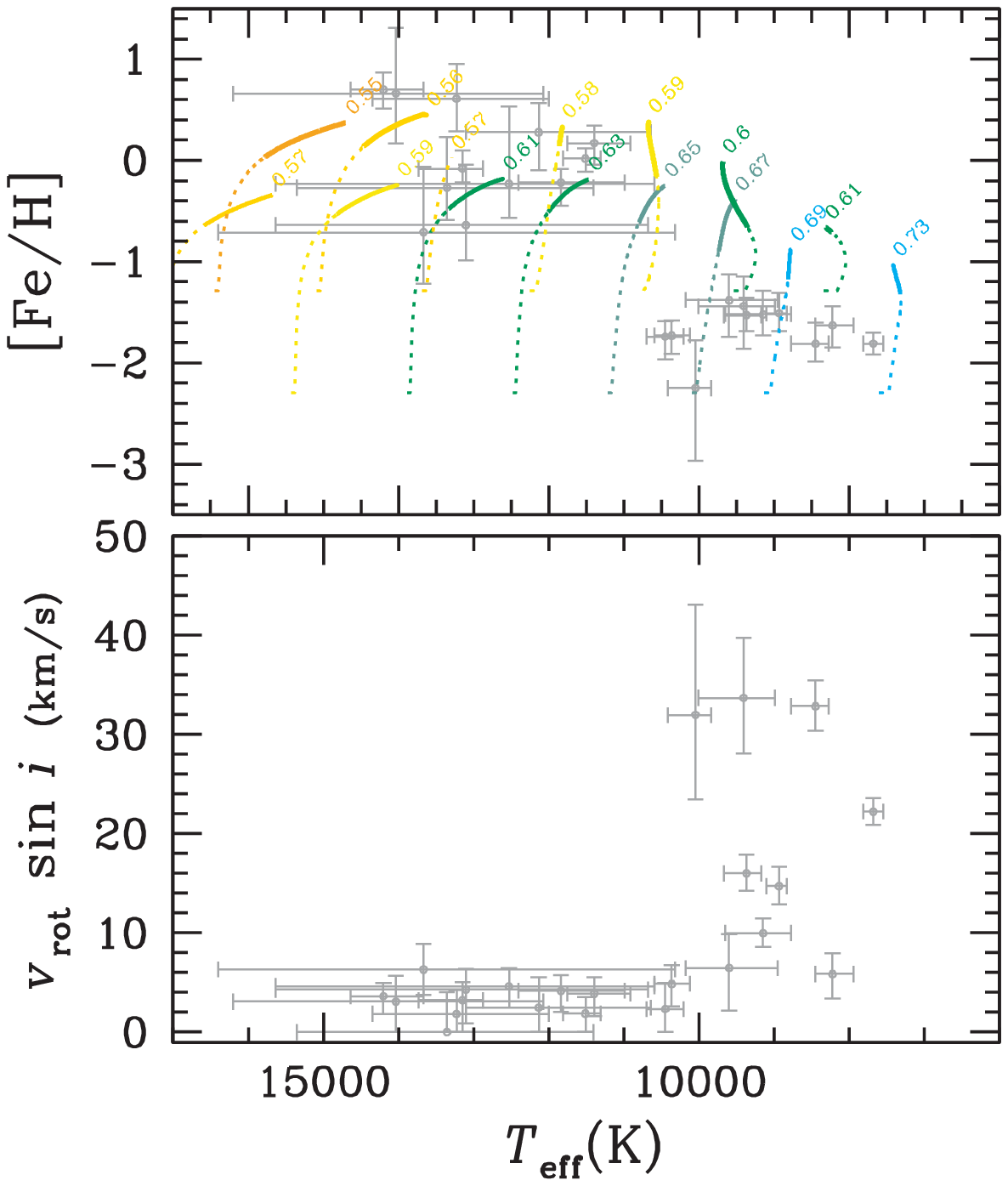}
\includegraphics[width=8cm]{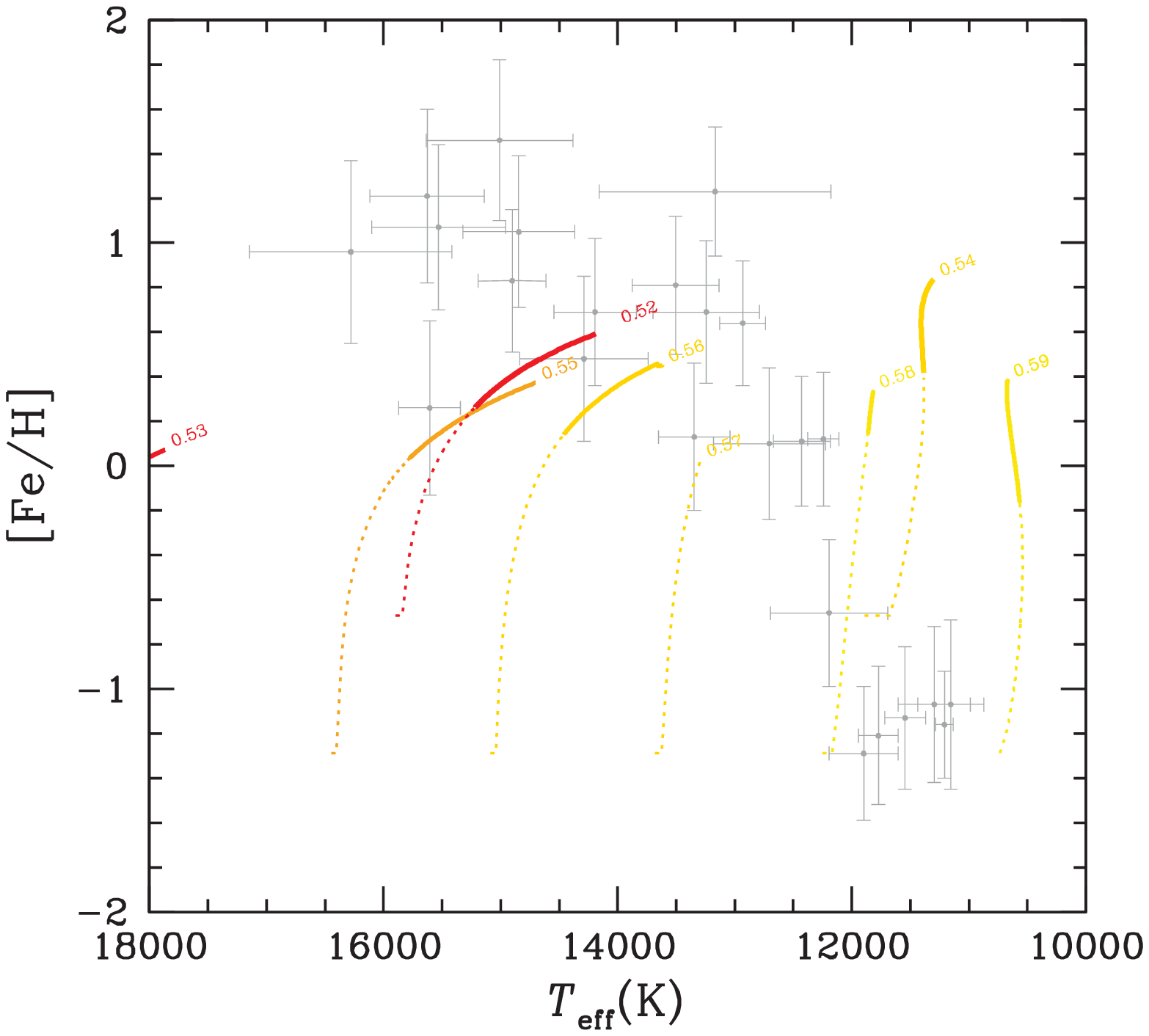}
      \caption{ \emph{Left panel} Surface concentration of Fe as a function of \teff{} in clusters (M3, M13 NGC288) from \citealt{Behr2003} with original Fe concentration between -2.3 and -1.3 dex the solar one. Models are shown for those two metallicities. Each  color coded segment represents the evolution of the surface Fe concentration. The line is solid for the time interval from 10 to 30 Myr after ZAHB but dotted from 0 to 10 Myr.  
\emph{Right panel} Surface concentrations of Fe as a function of 
      \teff{} in the cluster NGC\,2808 from \citealt{PaceRePietal2006}.
The  color coding is defined in footnote\,(\ref{foot:color}). For most
models the time interval spanned is approximately 30 Myr.            }
         \label{fig:NGC2808}
   \end{figure*}
  \begin{figure*}
   \centering
\includegraphics[width=9cm]{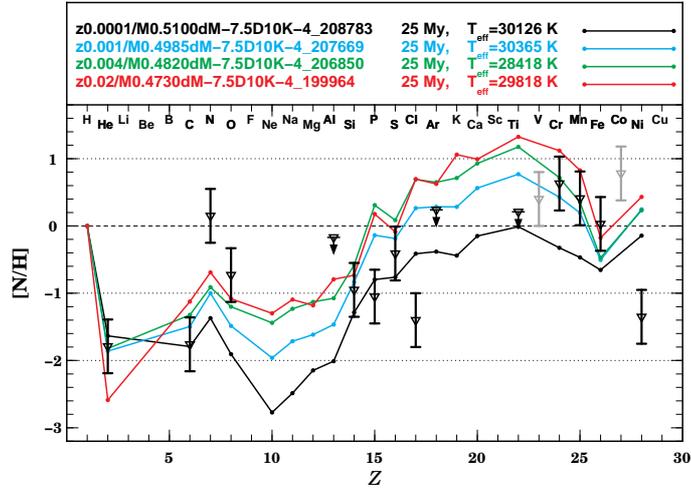}
      \caption{Surface abundances expected at 25\,Myr for  models  with turbulence anchored at  $\Delta M/\msol =10^{-7.5}$ calculated with $Z_0 = 0.0001$, 0.001, 0.004 and 0.02.    
      Observations are for PG0101+039 from \citet{BlanchetteChWeetal2008}. }.         
         \label{fig:blanchette001}
   \end{figure*}

\subsection{Intermediate metallicity globular clusters}
\label{sec:GlobularClusters}
 For intermediate metallicity clusters, comparisons are made in Fig.\,\ref{fig:M13} to patterns of anomalies for two stars in clusters\footnote{Variations in the abundances of Na and O, but not of Fe, as a function of \teff{} have recently been observed on the HB of M4 \citep{MarinoViMietal2010}.  However this is for cooler stars than the \Fe{} abundance variations discussed here and  in M4, the O and Na abundance variations  are probably due to nuclear reactions.  It would be of interest to verify the rotation velocities of the hotter stars in M4 given the correlations seen in Fig.\,\ref{fig:Fe0.0001}.}  (M\,13 and NGC\,1904) and in Fig.\,\ref{fig:NGC2808} for observations of Fe as a function of \teff{} in M\,3, M\,13, NGC\,288 and NGC\,2898.  
 
The expected patterns at six HB ages in a 0.55\msol{} model with original [Fe/H] = -1.3 are compared, in the left panel of Fig.\,\ref{fig:M13}, with WF4--3085 from the cluster M13 \citep{Behr2003}. According to Table\,1 of that paper, the cluster iron abundance is [Fe/H] = -1.54.  
In the right panel, comparisons are also made to models with  original [Fe/H] = -1.3 with star 469 of the cluster NGC1904 (M79)  \citep{FabbianReGretal2005} which has  $[\Fe/\H] = -1.59 $ according to \citet{KraftIv2003}. Since according to Fig.\,\ref{fig:2surfaces}, results are not too sensitive to metallicity (see also Fig.\,\ref{fig:blanchette001}) this comparison is accurate enough given that  models with original $[\Fe/\H] = -1.55 $ were not available.  In WF4--3085 the abundances are well reproduced by  models of 10\,Myr or more.  In 469, they are well reproduced by models of 5 Myr or more.  Given error bars,  all species heavier than Mg are well reproduced in both.  One may argue about P in WF4--3085 which is claimed to be overabundant by a factor of 500 while the model gives a factor of 50 overabundance.  The main difficulty is however with He which is observed to be underabundant by a much larger factor than models predict.  On the one hand this observation is probably difficult since these are relatively cool stars where the He abundance is more difficult to determine.  On the other hand, He is largey neutral in the atmosphere and this may enhance additional effets of diffusion there.  These will be further discussed in section\,\ref{sec:Results}.

In the right panel of Fig. \ref{fig:NGC2808}, the models with  [Fe/H] = -0.7 and [Fe/H] = -1.3 are compared to observations of NGC\,2808 from \citet{PaceRePietal2006}. The $[\Fe/\H]= -1.14$ determined by \citet{CarrettaBrCa2004} agrees within error bars with the $[\Fe/\H]$ of the cooler HB stars (below 12000\,K) and is bracketed by tracks for models with $[\Fe/\H] -1.3$ and -0.7. By zooming in on the figure, the origin (zero age HB) of each track may be clearly seen. The break between HB stars with the same Fe abundance as giants  and those with an overabundance by a factor of $\sim 20$ occurs slightly above 12000\,K which is $\sim 1000$\,K hotter than in the three clusters  M3, M13 and NGC288 in the left panel whose original [Fe/H] are bracketed by tracks for $[\Fe/\H] = -2.3$ and -1.3.  In Fig. 4 of \citet{QuievyChMietal2009}, the \teff{} interval from 10000 to 12000\,K is the interval over which anomalies are expected in only a fraction of the stars.  In M3, M13 and NGC288 (left panel), all stars with $\teff > 11000$\,K rotate slowly and have [Fe/H] within the range expected over most of the life time of a cluster with the original metallicity of the cluster.  The cooler stars have large enough rotation velocities for no anomalies to be expected\footnote{For an interesting correlation between [Fe/H],  \teff{} and rotation in HB and sdB stars  see Figure 8 of \citet{CortesSiReetal2009}.}.  Since we do not know the rotation velocities of the NGC\,2808 stars between 11000 and 12000\,K, we do not know if the 1000\,K difference in the \teff{} of the break in abundances is significant.  
One may then say that the run of \Fe{} abundances in BHB stars shown in Fig.\,\ref{fig:NGC2808} from 11000\,K to 17000\,K is well represented by calculations with a single value for the mixed mass of about $10^{-7}$\Msol.

 Finally the advantage of using clusters over field stars may be inferred from the right panel of Fig.\,\ref{fig:Fe0.0001}. There, the stars with $\teff < 11000$\,K give an indication of the original \Fe{} distribution in the field.  No information is however available on the original Fe abundance of individual stars with $\teff > 11000$\,K. So the link with rotation is more difficult to establish.  According to Fig.\,4 of \citet{QuievyChMietal2009}, the 13,500 and 15000\,K stars with rotation velocities of 30--40\,km\,s$^{-1}$ may or may not 
 rotate fast enough for atomic diffusion to be affected, depending on whether meridional circulation penetrates or not in the surface convection zone.  Since both stars might have a solar metallicity to start with, little can be inferred from their relatively large rotation velocities.  The other field stars with large metallicity all have very small rotation velocities as was the case in clusters.
   \begin{figure*}
   \centering 
\includegraphics[width=10cm]{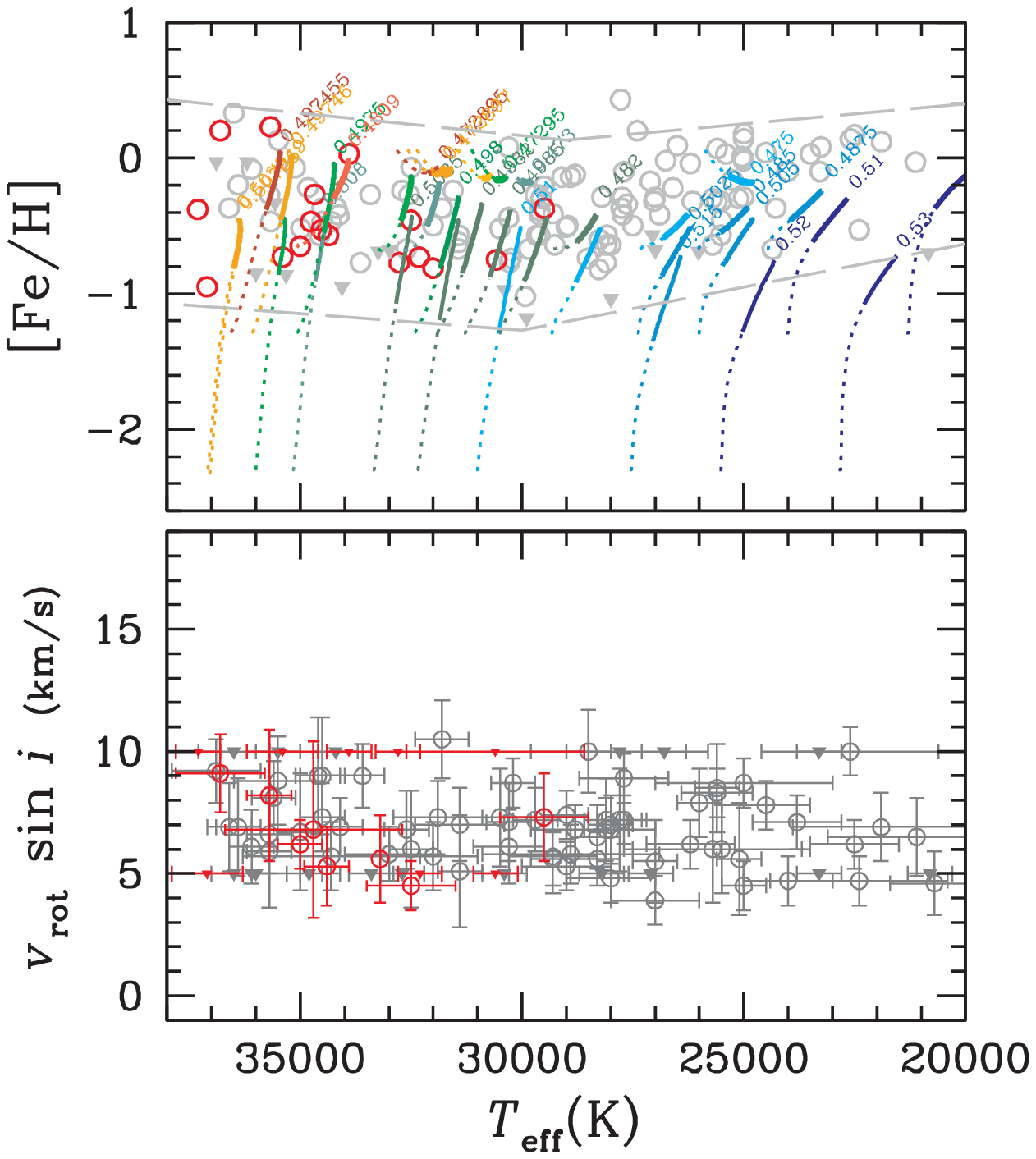}
      \caption{   Surface concentrations of Fe as a function of       \teff{} in sdB stars.
Models were calculated with original concentrations of [Fe/H]$ = -2.3, -1.3, -0.4$ and 0.0. Models span the \teff{} interval
from 20\,000 to  37\,000\,K   and the mass interval from 0.47 to 0.52\,\Msol{}.
The lines give the surface abundance variation for each calculated model.  They are color coded as a function of  the mass of the model which may be read by zooming on the figure in the electronic version; solid lines for the time interval from 10 to 32 Myr after ZAHB and dotted
 from 0 to 10 Myr. The dotted line starts at the original abundance of the model.  For most
models the time interval spanned is approximately 32 Myr.  Note that most models starting with a solar metallicity have very short tracks.
      Observations (circles) are from \citealt{GeierHeNa2008}, \citealt{GeierHeEdetal2010} and Geier\,2010, private communication.  Inverted triangles are upper limits. The stars in red in Fig.\,\ref{fig:DMfixed} are also in red in this figure.    The bottom part of the figure gives the rotation velocities and the error bars on \teff.   
       }
         \label{fig:Fe-sdb}
   \end{figure*}
 
  \begin{figure*}
   \centering
\includegraphics[width=16cm]{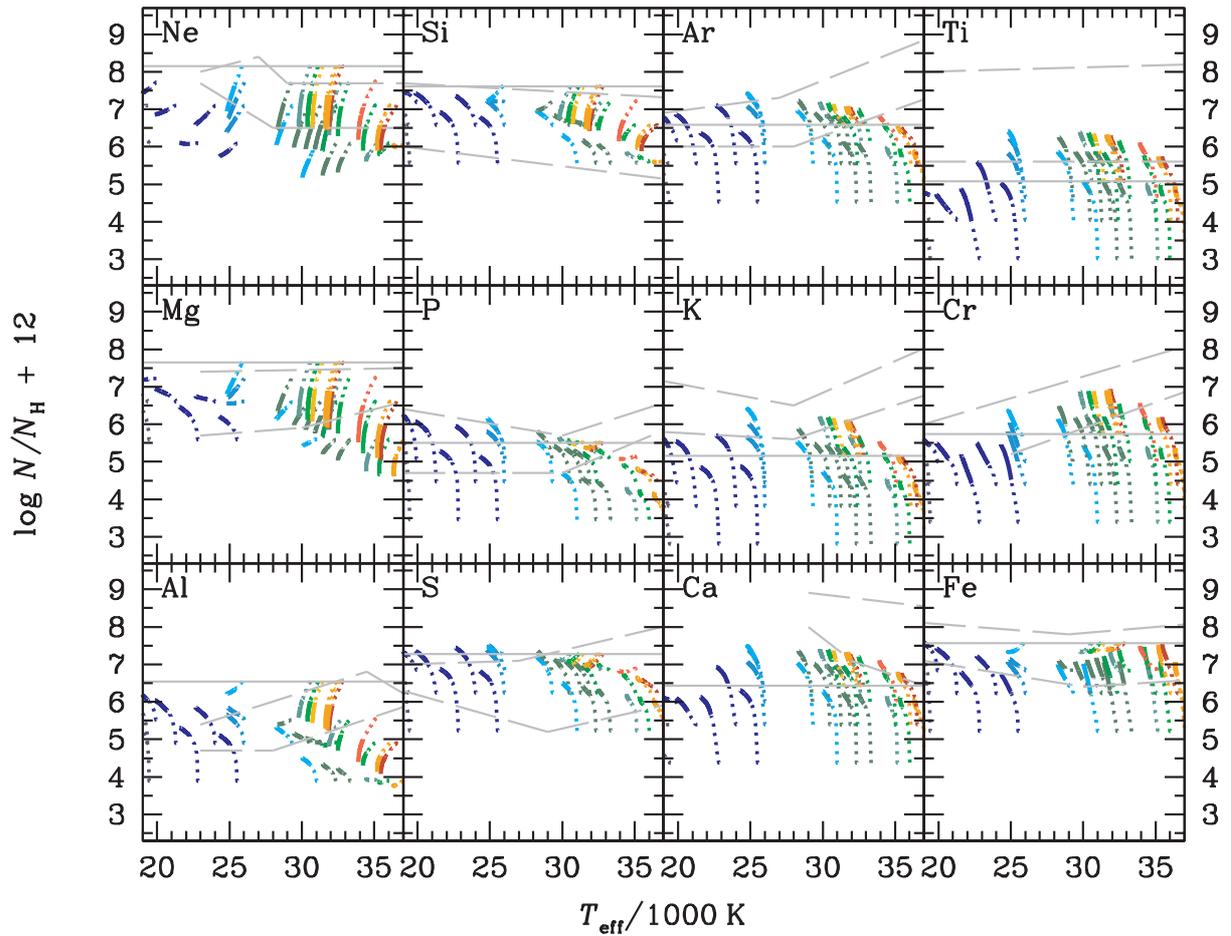}
      \caption{Surface abundances for atomic species from Ne to Fe.  Iron  is also on Fig.\,\ref{fig:Fe-sdb} where  is found a detailed caption.  The domain of observed values is bounded  by long dashed broken gray lines which are also shown for Fe on  Fig.\,\ref{fig:Fe-sdb}. The domain of abundances often extends below the lower curve since many measurements are upper limits (see \citealt{GeierHeNa2008} and \citealt{GeierHeEdetal2010}).  The solid horizontal grey line is the solar abundance.   Note that individual panels span 7 orders of magnitude in abundances. }
         \label{fig:heber001}
   \end{figure*}
    
    \begin{figure*}
   \centering 
\includegraphics[width=10cm]{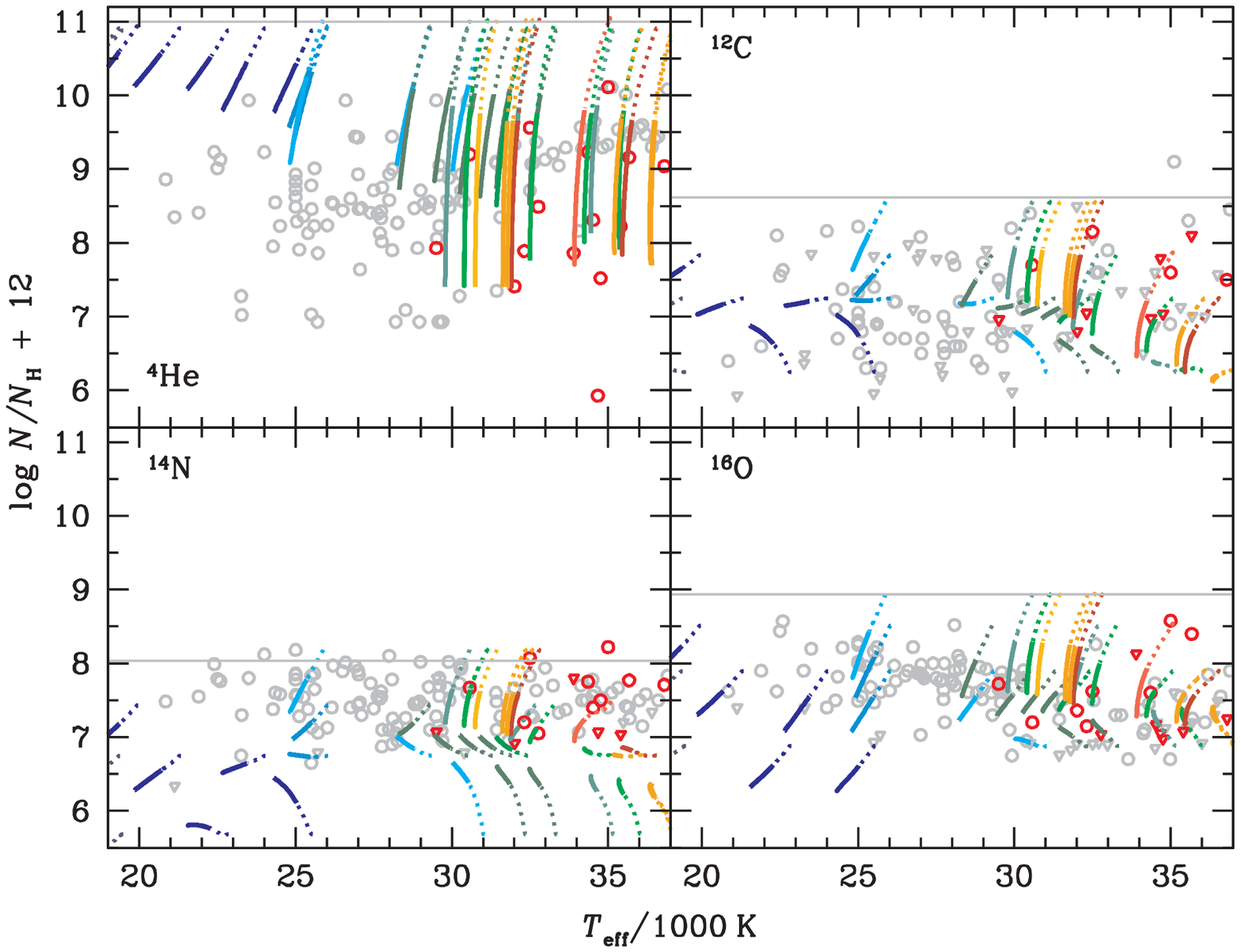}
      \caption{   Surface concentration of He as a function of 
      \teff{} in sdB stars. See the caption of Fig.\,\ref{fig:Fe-sdb} for the definition of the curves.
      Observations (circles) are from \citealt{GeierHeNa2008}, \citealt{GeierHeEdetal2010} and Geier (2010, private communication).  The stars in red in Fig.\,\ref{fig:DMfixed} are also in red in this figure.   
The lines for He were stopped between $X(\He) = 0.001$ and 0.0001 because the evaluation of the surface \He{} abundance becomes inaccurate for smaller values.   }
         \label{fig:He-sdb}
   \end{figure*}

\subsection{Field sdB stars}
\label{sec:FieldSdBStars}
 For sdB stars, neither the age nor the metallicity are known.  Anomaly patterns are compared to observations for two stars with models of four different original metallicities.
Comparisons are made with sdBs of $\teff{} \sim 31000$\,K using data from \citet{OtooleHe2006} (Feige 48, see right panel of Fig.\,\ref{fig:2surfaces}) and $\teff{} \sim 30000$\,K  in Fig.\,\ref{fig:blanchette001} using data from \citet{BlanchetteChWeetal2008} for PG0101+039.    We decided to compare to models of 25\,Myr, whose surface abundances are representative of a significant fraction of the evolutionary span.  For Feige 48 (Fig.\,\ref{fig:2surfaces}), 14 of the 17 species which were both calculated (with original [Fe/H] between $-0.7$ and $-1.3$ or $Z_0$ between 0.001 and 0.004) and observed agree reasonnably well. For PG0101+039 (Fig.\,\ref{fig:blanchette001}), 15 species are both calculated and observed and 12 are in reasonable agreement with either of the two lower metallicity models.  The agreement is however  not perfect but this is perhaps not too surprising given the uncertainty of the age.

Once neither age nor original metallicity are known, perhaps as useful information is obtained by comparing an ensemble of models spanning the metallicity, mass and age intervals of  a large number of stars.  Models of four metallicities are used with ages spanning the first  32 Myr of HB evolution and a mass interval leading to the observed \teff{} interval of 20000 to 37000\,K. This \teff{} interval corresponds to that of a recent survey 
 of the chemical composition of sdB stars whose preliminary results are published in \citet{GeierHeNa2008} and \citet{GeierHeEdetal2010} with additional private communications to us from Dr Geier.
On Fig.\,\ref{fig:Fe-sdb}, observations of $X(\Fe)$ are the grey (or red) circles with triangles being upper limits.  The long dashed grey lines define the interval where observations are found.  They are similarly defined for other elements from the observations and are shown in order to constrain the observed abundance interval in Fig.\,\ref{fig:heber001} and \ref{fig:He-sdb}.  The rotation velocities are shown on the lower panel of Fig.\,\ref{fig:Fe-sdb}.  They are all smaller than 10\,km/s which is too small to interfere with atomic diffusion at those \teff{} according to \citet{QuievyChMietal2009}.  The uncertainty on \teff{} is also shown in that panel.
 
Comparisons are made as a function of \teff{} using observations from \citet{GeierHeNa2008} and \citet{GeierHeEdetal2010} in Fig.\,\ref{fig:heber001} for Ne, Mg, Al, Si, P, S, Ar, K, Ca,Ti, Cr and Fe and in Fig.\,\ref{fig:He-sdb} for He and CNO. These are  the species  for which observations are reported by \citet{GeierHeEdetal2010} and which are also present in the OPAL opacity calculations.  An indication of the limits of the domain of observed points is given by the long dashed gray lines which are defined for Fe in Fig.\,\ref{fig:Fe-sdb}.  For most atomic species, the lower broken gray line corresponds to as many upper limits as data points.  We refer the reader to the original publication of the data.

In all three  figures, for each model and each chemical species the abundance is indicated by a curve originating at the original abundance, dotted for the first 10\,Myr and solid until the end of the calculation, usually around 32 Myr.  The various colors  distinguish masses (see footnote\,{\ref{foot:color}}).  The masses are indicated at the end of each colored curve on Fig.\,\ref{fig:Fe-sdb} and may be read more easily by zooming in the electronic version. In doing the comparison between observations and calculations, one should take into account that the calculations covered the first 32 Myr of the HB evolution. Since HB evolution probably lasts three times longer the colored lines should often be extended by a factor of 2, or perhaps 3 when no saturation is involved. When saturation is important the lines can be much shorter however as is seen for many models which originally had the solar abundance (these are easily identified since their dotted line starts at the solid horizontal  gray line in Fig.\,\ref{fig:heber001}).  

For Fe (Fig.\,\ref{fig:Fe-sdb}), the calculated values cover the range of observed Fe abundances at all \teff's especially when the expected extension of the curves to the end of the HB evolution is taken into account.  The concentration of Fe abundances between solar and 10 times below solar is to be expected given the expected original distribution of Fe abundances as seen in the right panel of Fig.\,\ref{fig:Fe0.0001}.  A number of upper limits have been measured (the grey triangles) and can be accounted for  by the early evolution of stars starting with very small original Fe abundances.  Even those, however, end up spending most of their HB evolution within the main interval of the observations.  The objects represented by red circles are also compatible with the results presented here even though they have a smaller gravity, and so are probably more evolved than the models (see Fig.\,\ref{fig:DMfixed}).

The comparison between calculated and observed abundances for elements from Ne to Fe is shown on Fig.\,\ref{fig:heber001}.  Iron is repeated to allow evaluating the effect of representing observations as lying between the two long dashed gray lines of each panel.  However the lower bound frequently appears to correspond to the smallest abundance that can be detected since it  generally corresponds to upper limits by the data points which may be found in \citet{GeierHeNa2008} and \citet{GeierHeEdetal2010}.  The reader will find useful looking at the original data.  In two cases, Si and S, observed abundances cover the same interval as the simulations' starting abundances (the dotted lines originate within or very close to the gray lines).  In both cases they remain within those bounds.  For Si, the small observed \teff{} dependence is obtained while for S, the observed slight increase around 35000\,K is not reproduced.  In four cases (K, Ca, Ti and Cr), the observed abundances are 4 to 5 orders of magnitude larger than the original abundances in the more metal poor original models. Taking into consideration the number of upper limits in the observations, and that only the first third of HB evolution is covered by the lines, the agreement is generally good.  One property which is not reproduced is the apparent increase in observed abundance above 33000\,K for K and Cr.  For P and Ar the observed values are compatible with observations except for $\teff > 33000$\,K; however for P there  are mainly upper limits.

For C and N, (see Fig.\,\ref{fig:He-sdb}) there seems to be strong disagreement for  $\teff > 33000$\,K while at lower \teff, there is agreement for C while N is on average perhaps $3\times$ more abundant than expected.  This  could be related to some nuclear effects.
On the other hand calculated  O seems generally compatible with observations.

Our calculations of the surface He abundance compared to observations on Fig.\,\ref{fig:He-sdb}, are constrained by a limitation of our code since in our calculation algorithm, $X(\He)$ is obtained by subtraction\footnote{It is the only element for which this is done.  Eliminating this limitation requires rewriting the central part of the evolutionary code involving the transformation of H into He.}. Consequently the lines for He were stopped between $X(\He) = 0.001$ and 0.0001 because the evaluation of the surface \He{} abundance becomes inaccurate for smaller values, when $Y \ll Z$.   
One has the surprising result that observations of $X(\He)$ are in agreement with calculations in the stars with $\teff > 29000$\,K but not cooler.  The agreement for stars with $\teff \sim 30000$\,K is reinforced by the detailed comparisons shown on Figs.\,\ref{fig:2surfaces} and \ref{fig:blanchette001}. In the higher \teff{} stars, the observed abundance range is easily covered once  one takes into account that calculations cover only the first third of the HB life span.  Whilst still helping at lower \teff{}, this is not sufficient to explain the absence of stars with an underabundance smaller than a factor of 0.1 nor the presence of stars underabundant by a factor of $10^{-4}$.  This reinforces the problem with He abundance in globular cluster stars (see Fig.\,\ref{fig:M13}) and is discussed in Sect.\,\ref{sec:Results}.


\section{Conclusion and General Discussion}
\label{sec:Conclusion}

Stellar models have been evolved over the first 32\,Myrs of HB evolution for masses leading to the \teff{} interval of 8000 to 37000\,K and for metallicities ranging from $Z_0 = 10^{-4}$ to 0.02. A  total of some 60 models were evolved.  They all started from the evolution, from the zero age \MS{} to the He flash, of 0.8\ to 1.0\,\msol{} models of metallicities from $Z_0 = 10^{-4}$ to 0.02 described in \citet{MichaudRiRi2010} (see in particular their Sect. 2).  They corrrespond to the metallicities of the various globular  clusters whose horizontal branch has been studied.  They also cover the metallicity (right panel of Fig.\,\ref{fig:Fe0.0001}) and \teff{} intervals (right panel of Fig.\,\ref{fig:DMfixed}) of field sdB stars. 

\subsection{The main results}
\label{sec:Results}
One remarkable observational property of both field sdB stars (studied here up to 37000\, K) and cluster HB stars with $\teff > 11000$\,K is that their Fe abundance is relatively close to solar whatever the metallicity of the cluster and independent of the unknown original Fe abundance in the case of sdB stars.  In the calculations described above for field stars, the original metallicity was varied from $Z_0 = 0.0001$ to $Z_0 = 0.02$.  In globular clusters, the original abundances for the calculations are those of giants of the cluster. So long as HB stars have their outer 10$^{-7}$\msol{} mixed, it has been shown to follow naturally from the stellar evolution with atomic diffusion  that the final Fe abundance ended within a factor of ten interval nearly independent of \teff{} (see Sec.\,\ref{sec:Comparison} and in particular Figures \ref{fig:Fe0.0001}, \ref{fig:NGC2808} and \ref{fig:Fe-sdb}). Overabundances of Fe by factors of up to 100 are implied. Reducing the mixed mass by a factor of ten or more leads to unacceptably large Fe abundances (see Sec.\,\ref{sec:MixedMass} and Fig.\,\ref{fig:Fe_gr}).   The well observed Fe abundance so constrains the mixed zone in stellar evolution models\footnote{If, instead of using Eq.\,(\ref{eq:Delta-M-0}) one used Eq.\,(\ref{eq:rho-T-0}) so that  turbulence were fixed at a given $T_0$, the mixed mass would vary by more than 3 orders of magnitude  between  stars of  $ \teff{}=12000$ and  30000\,K. We have verified that starting with $Z_{\rm{0}} = 0.001$, a 12000 K HB star would develop solar surface \Fe{} abundance while a 30000\,K star would develop a Fe abundance 50 times solar.
This would lead to completely unacceptable changes in relative surface abundances when taking sdBs and HB stars of clusters into account (see Figs.\, \ref{fig:NGC2808} and \ref{fig:Fe-sdb} for observed surface \Fe{} abundances and also Fig.\,\ref{fig:Fe_gr} for the effect of a change of the mixed mass by 2.5 orders of magnitude).}.  Is this mixing compatible with observation of other atomic species?

Detailed comparisons were carried out for a number of sdBs and cluster stars in which the abundance of many elements were determined.  In all cases the comparison between observations and model calculations led to acceptable results (see Sec.\,\ref{sec:Comparison} and Figs.\,\ref{fig:2surfaces}, \ref{fig:M15}, \ref{fig:M13} and \ref{fig:blanchette001}). 
Approximately 12 of 15 elements are in reasonable agreement in each case.  This is obtained without any adjustment since the only adjustable parameter was fixed by the Fe abundance.  This is true in globular clusters whose original [Fe/H] varies from --2.3 to --0.7 as well as in field sdB stars.  For these, a comparison to a large sample of field stars covering a large age and original metallicity interval was also   carried out with general agreement (see Figs.\,\ref{fig:heber001} and \ref{fig:He-sdb}).

The calculated surface abundance of He  agrees with that observed in the sdB stars shown in Figs. \ref{fig:2surfaces} and \ref{fig:blanchette001} and also for $\teff > 25000$\,K in Fig.\,\ref{fig:He-sdb} but there appears to be a large discrepancy in the cooler stars and in particular those HB stars shown in Fig.\,\ref{fig:M13} and the left panel of Fig.\,\ref{fig:2surfaces}.  In the latter cases, \teff{} is small enough that He is not ionized in the atmosphere.  The neutral He atomic diffusion coefficient is  $\sim \times 100$ larger than the ionized one \citep{MichaudMaRa78}.  Additional element separation could be occuring in the stellar atmosphere  for He since its settling velocity there would be enhanced by its being neutral. There is observational evidence for such stratification of \Fe{} \citep{KhalackLeBoetal2007,KhalackLeBe2010,LeBlancMoHuetal2009,LeblancHuKh2010}.  Other species, for which \gr{} in the atmosphere would happen to be very large,  could also be affected by separation in the atmosphere if the mixing there is not very strong. 
The evolutionary models described in this paper may then not tell the whole story.  In real stars, there  could be additional  abundance variations between the surface and the bottom of the mixed mass of some $10^{-7} \Msol{}$. 
  
Since in particular gravity varies by two orders of magnitude over that \teff{} interval on the HB, the similarity of observed abundances is quite surprising.  It is clearly linked to the saturation of radiative accelerations.

The agreement between the observations over the whole HB and the expected abundance anomalies from the evolution calculations leaves little doubt that radiative accelerations are the main cause of abundance anomalies on the HB.   

\subsection{Turbulence or mass loss}
\label{sec:OrMassLoss}
In  this paper it was  found  that if the outer $10^{-7} \Msol{}$ was mixed\footnote{The first mesh point in the models of this paper was typically at $10^{-15}\Msol{}$.} by turbulence most observed abundance anomalies followed. However as soon as an adjustable parameter has an effect on the results, one must remain cautious and, in spite of the observational support described above, one must question if the physical process assumed to do the mixing is the only one possible or if there exist alternatives.  In the only stellar evolution calculations we know of where mixing (or homogeneity) was not assumed for the outer region,  
it has been shown by \citet{VickMiRietal2010} that abundance anomalies observed on AmFm stars could be reproduced as well by a model that assumes mass loss as by the model where turbulence is in competition with atomic diffusion \citep{RicherMiTu2000,TalonRiMi2006}.  They found that the velocity corresponding to the mass loss rate had to equal the atomic diffusion velocity at the stellar mass fraction where the mixed zone ends in the turbulence model.  This occurs here at $10^{-7} \Msol{}$.  Presumably the same would be true in HB stars.  Instead of turbulence, mass loss could play a role to reduce anomalies.  One may evaluate the required mass loss rate by equating the corresponding velocity to the He settling velocity at $10^{-7} \Msol{}$ using the expression  
\begin{equation}
	\frac{dM}{dt} = -4 \pi r^2 \rho v_{\rm{drift}}(\He).
	\label{eq:mass_loss}
\end{equation}
 It was found to equal 3--5 10$^{-14}$\Msol/yr at $\DM = -7$,  in a model with $\teff \sim 30000$\,K. A ten times  smaller  mass loss rate has been suggested before to explain in particular observations of Si in sdB stars \citep{MichaudBeWeetal85}.  The calculations included only the outer $10^{-5} \Msol{}$ and the separation occured right in the atmosphere.  Both the mass loss rates and the timescales of the calculations were smaller. A  complete stellar evolution calculation including mass loss instead of turbulence as the mixing process seems justified for HB stars,  along the lines done by \citet{VickMiRietal2010} for \MS{} stars.

\subsection{$\mu$ gradient inversions}
\label{sec:Mu}
It was shown in Sect.\,\ref{sec:MixedMass} that when the mixed mass is smaller than $10^{-7} \Msol{}$ an inversion of the mean molecular weight often occurs (see Fig.\,\ref{fig:Fe_gr}).
It has been suggested in a similar  context by \citet{TheadoVaAletal2009} that such a $\mu_0$ inversion is unstable and should lead to mixing.  It is tempting to conclude that this $\mu_0$ inversion is \emph{the} cause of the mixing implied by observations. However their analysis does not include the effect of \gr{} on the instability.  Since the radiative accelerations are the cause of the $\mu_0$ inversion, their role  in the  analysis of the hydrodynamical instability could be important. The metals which are supported by \gr{}s do not contribute to the increase of the weight of the material.  So one may wish to consider an effective $\mu_0$ for which supported metals contribute negatively.
Intuitively, one may assume that if the function
\begin{equation}
\mu_{0\mathrm{eff}} = \frac{\sum N(A_i) A_i m_p [g-\gr(A_i)]} {\sum N(A_i) m_p g}
	\label{eq:muoeff}
\end{equation}
does not decrease inwards where $\mu_0$ decreases inwards because of \gr{}s (which cause some metal abundances to increase outwards) then \gr{}s may maintain stability.  This function $\mu_{0\mathrm{eff}}$ is shown (inverted) by the dotted curves (see Fig.\,\ref{fig:Fe_gr}).  Contrary to $1/\mu_0$, $1/\mu_{0\mathrm{eff}}$ increases outwards where the Fe and Ni abundances increase outward.  It seems to us that one should not conclude that a $\mu_0$ inversion caused by \gr{}s on metals leads immediately to an instability;  further work is needed to investigate the nature of the $\mu{}$ inversion instability in the presence of radiative accelerations.

\subsection{Asterosismology}
\label{sec:Asterosismology}
In their ground breaking  work, \citet{CharpinetFoBretal96}   have established that the \Fe{} abundance expected to follow from atomic diffusion in the envelope of sdB stars should lead to pulsations. This was confirmed observationnally \citep{KilkennyKoOdetal97}.  See also  the reviews  of \citet{CharpinetFoBr2001}  and \citet{FontaineBrChetal2008}.  Their approach has the advantage to involve no adjustable parameter since they assume equilibrium between \gr{} and gravity in the driving region for the pulsations.  This implies that mass loss and turbulence  have a negligible effect.  It is however not established if in sdBs with small original metallicity  there is enough Fe to fill sufficiently the region where the driving occurs.  As noted in the conclusion to the tenth sdB for which a detailed asterosismic analysis had yielded the mass \citep{RandallVaFoetal2009}  a more accurate fit to the asterosismic data requires evolutionary models of sdB stars.  
The approach to equilibrium in the driving region was studied by \citet{FontaineBrChetal2006} starting from a \emph{solar} Fe abundance. They  showed using static models that the Fe abundance became able to drive pulsations within $10^5$\,yrs.  These works are based on a detailed analysis of the structure of the outer regions of these stars.

Our results cleary show that from whatever Fe abundance sdB stars start with, a sdB star model with the outer 10$^{-7}$\msol{}  mixed ends up  with a solar Fe abundance in a fraction of the HB lifetime of 100 Myr.  
This could  be used as a starting abundance for the calculations of  \citet{FontaineBrChetal2006} mentioned above.  In fact, the curve 
labeled $-9$ in Fig.\,\ref{fig:Fe_gr} shows that after 32\,Myr, starting from a metallicity 20 times below solar, the complete evolutionary models calculated here with OPAL data have accumulated approximately the same \Fe{} abundance
as shown in Fig.\,1 of \citet{FontaineBrChetal2006} at the same age for a model starting with a solar abundance.  According to Fig.\,\ref{fig:Fe-sdb},  models starting with an Fe abundance 200 times below solar end up with 3 times less \Fe{} after 32\,Myr but according to Fig.\,1 and 3 of \citet{FontaineBrChetal2006} this is still sufficient to excite the spectrum.

More recently, using  an analysis of the Fe abundance required to cause observed pulsations in sdBs, \citet{CharpinetFoBr2009} conclude that if Fe is the sole atomic species responsible for the pulsation of sdB stars, the Fe abundance needs to be $\log N_{\Fe}/N_{\H} \sim -4.09$ and $-3.75$ at $ \teff = 29580$ and 35050\,K respectively  in the driving region.  This is larger than the observed \Fe{} abundance and suggests that some separation needs to occur between the driving region and the surface.  However from Fig.\,\ref{fig:interiorX}, \Ni{} is easily a factor of 3 more overabundant than \Fe{} and it may contribute significantly to driving the pulsations as suggested by \citet{JefferySa2007} and just as it contributes significantly to iron convection zones in Pop.I stars (see Fig.\,4 of \citealt{RichardMiRi2001}).

The models described in this paper clearly need to be tested by calculating their pulsation spectra.  When models with mass loss become available (see Sect.\,\ref{sec:OrMassLoss}), asterosismology could help decide which is the main  mechanism competing with atomic diffusion driven by \gr{}'s.  This is however outside  the scope of the present paper.

\begin{acknowledgements}
 We thank Drs Ulrich Heber and Stephan Geier  for very kindly communicating to us their data in tabular form.   We thank an anonymous referee for  a very careful reading of the manuscript and for numerous constructive remarks. This research was partially supported at  the Universit\'e de Montr\'eal 
by NSERC. We thank the R\'eseau qu\'eb\'ecois de calcul de haute
performance (RQCHP)
for providing us with the computational resources required for this
work.

\end{acknowledgements}
\bibliography{astrojabb,michaud}
\bibliographystyle{aa}
%
\Online

\begin{appendix} 
\section{Black and white version of Figure\,\ref{fig:Eventails}}
      \begin{figure*}
   \centering
\includegraphics[width=9cm]{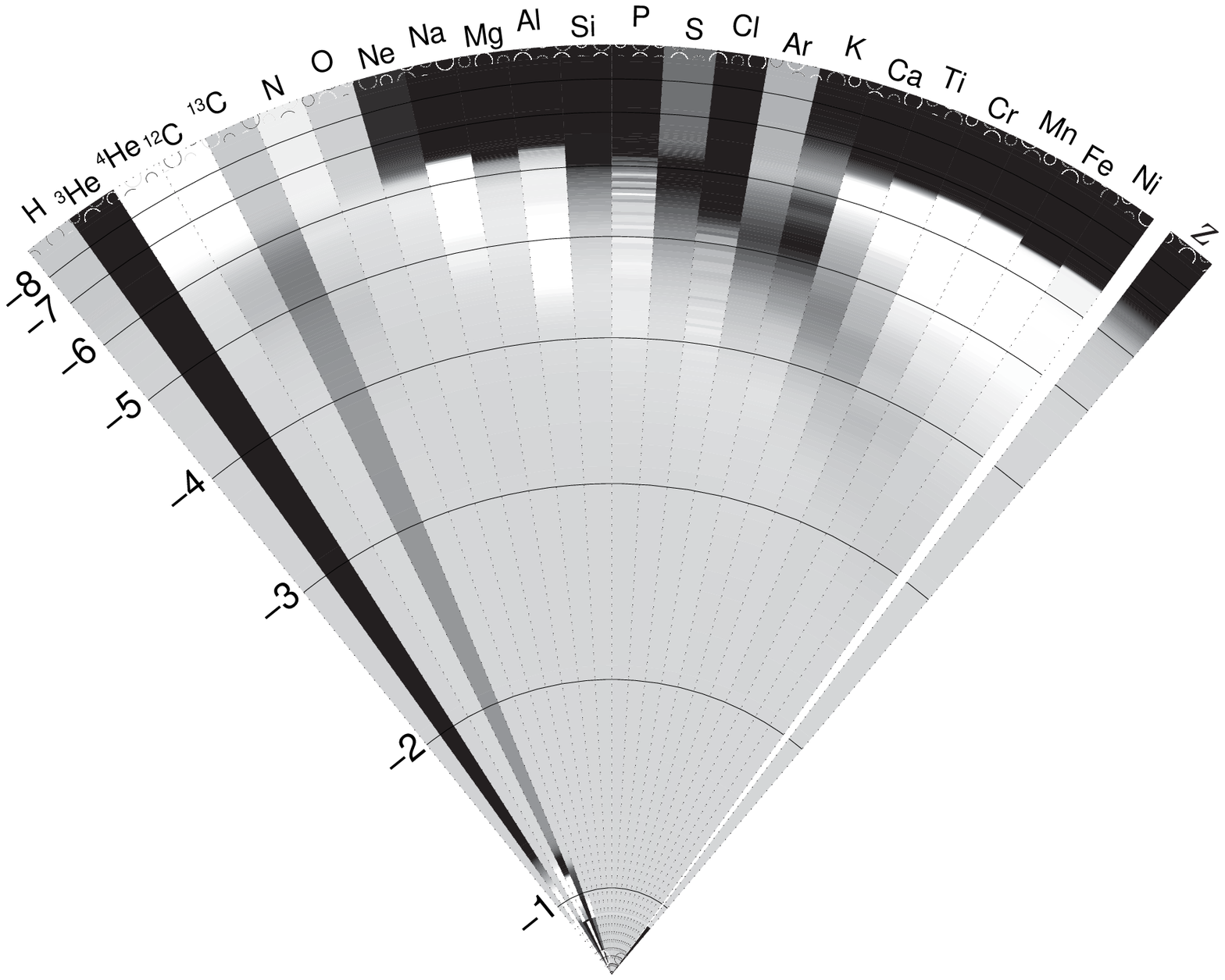}
\includegraphics[width=9cm]{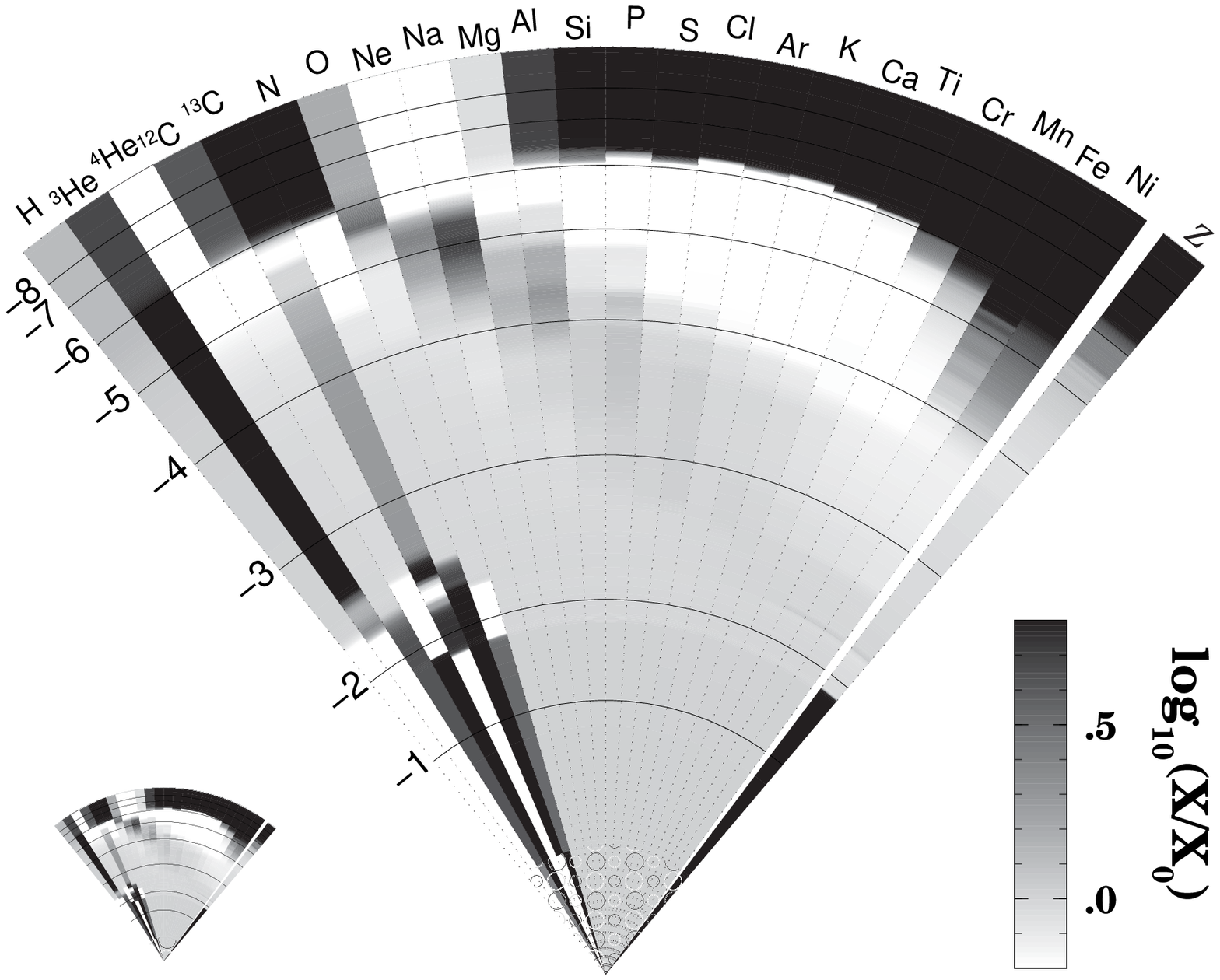}
      \caption{Black and white version of Fig.\,\ref{fig:Eventails}.  Gray coded concentrations in two HB stars of the same metallicity after 25\,Myr on the HB. Left panel with a \teff{} of  14\,000\,K (0.59\,\msol{}) and right panel of 30\,100\,K (0.51\,\msol).  The scale of the 	radius is linear but the logarithmic value of the mass coordinate above a number of points, \DM, is shown on the left of the 
	horizontal black line.  The concentration scale is given in the right insert.  
	  Small circles near the top of the left panel
	mark the extent of the surface convection zone while similar circles near the center of both models mark the central convection zone.   The small inset in between the two panels shows the high \teff{} star on the radius scale of the low \teff{} star.    For $-7<\DM < -4$ the concentration is quite different for many species.  It is surprisingly so for C and O for $\DM>-2$.    }
         \label{fig:EventailsBW}
   \end{figure*} 

\clearpage
\section{Properties of evolutionary models}
\label{sec:Appendix}

\begin{figure*}
   \centering
\includegraphics[width=16cm]{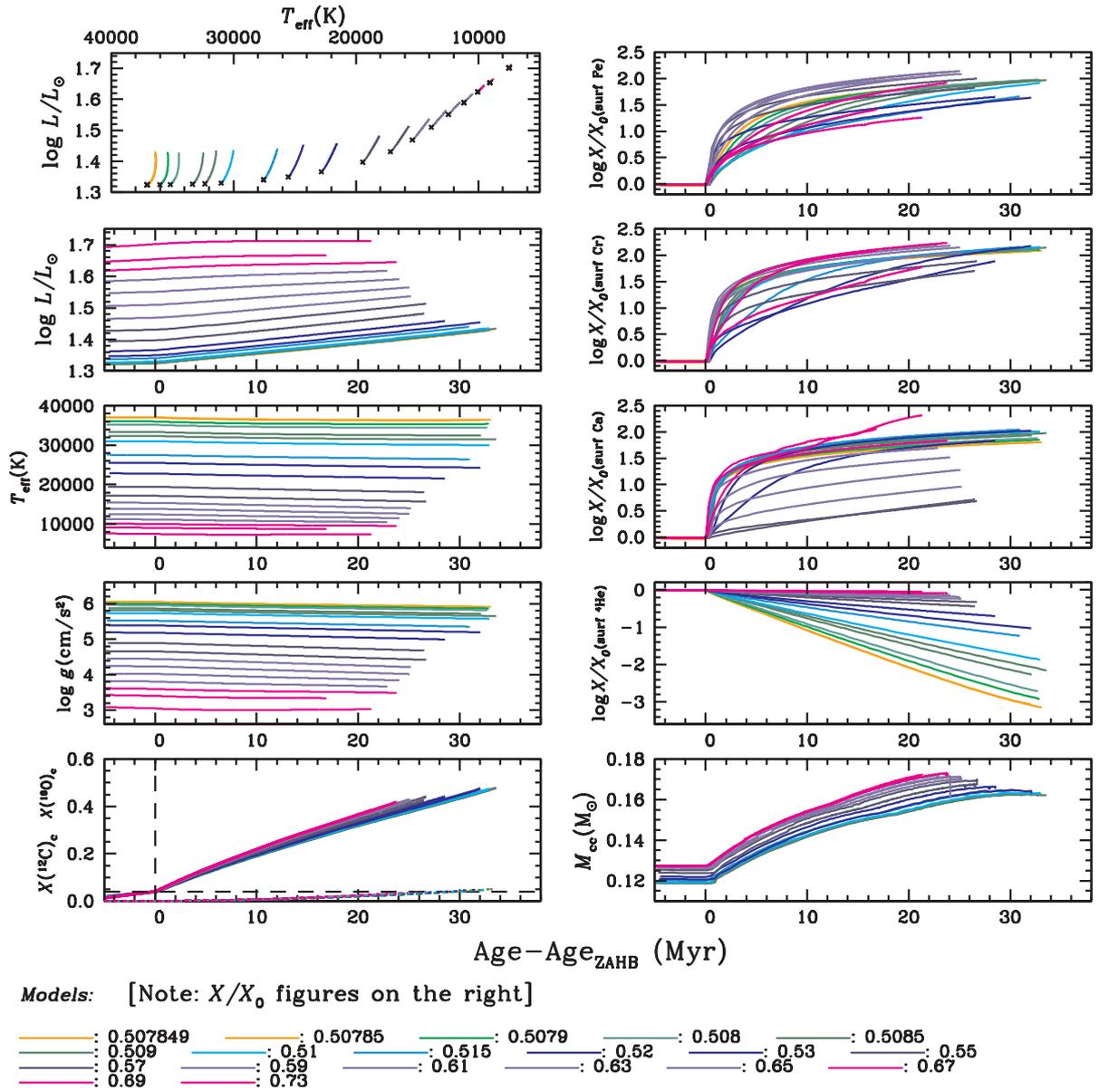}
      \caption{Models with $Z_0 = 0.0001$. The curves are identified on the figure by the mass of the HB model.   The color code is the same as used for the left panel of Fig.\,\ref{fig:DMfixed} and is defined in footnote\, {[\ref{foot:color}]}. 
      Zero age HB is defined as the moment when $X(^{12}\mbox{C})$ reaches 0.04.  On the figures the curves are not all very precisely aligned to that value.  $X_0(\mbox{surf Fe})$ identifies the \Fe{} mass fraction in the original \MS{} model. Similarly for the other species.  
      }
         \label{fig:modelsZ0001}
   \end{figure*}

\begin{figure*}
   \centering
\includegraphics[width=16cm]{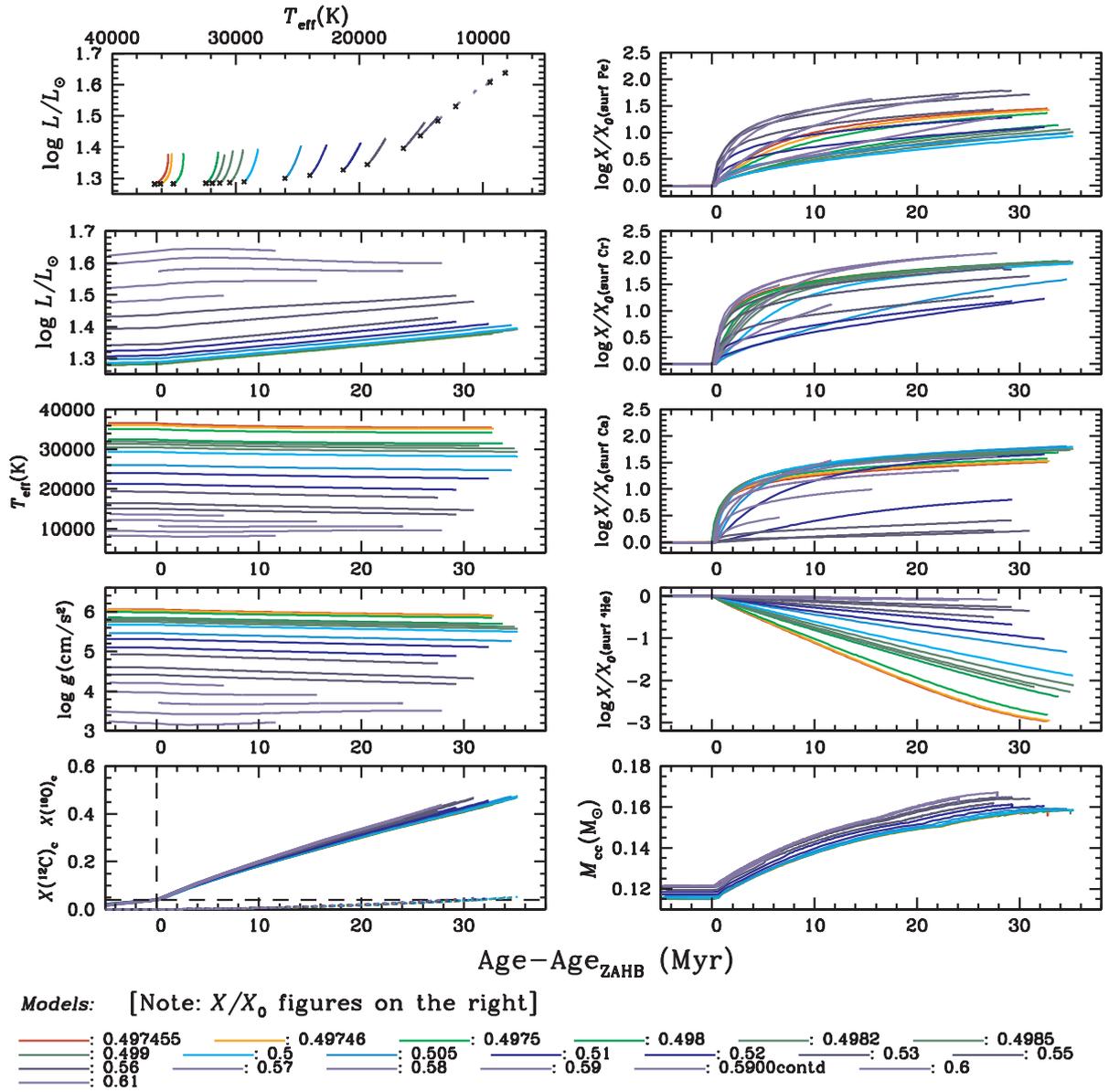}
      \caption{ Models with $Z_0 = 0.001$.  See caption of Fig.\,\ref{fig:modelsZ0001}.}
         \label{fig:modelsZ001}
   \end{figure*}

\begin{figure*}
   \centering
\includegraphics[width=16cm]{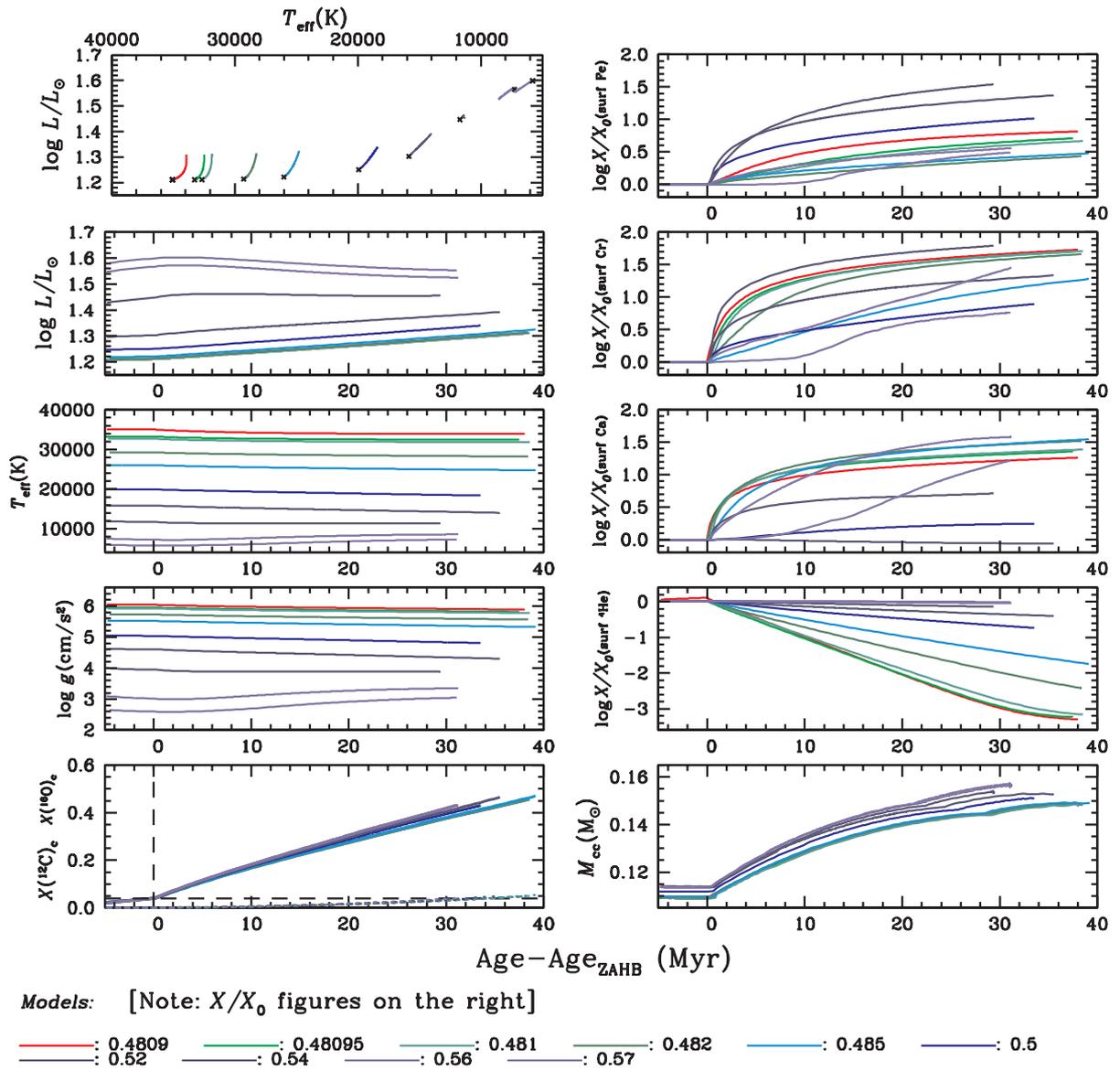}
      \caption{Models with $Z_0 = 0.004$. See caption of Fig.\,\ref{fig:modelsZ0001}.}
         \label{fig:modelsZ004}
   \end{figure*}

\begin{figure*}
   \centering
\includegraphics[width=16cm]{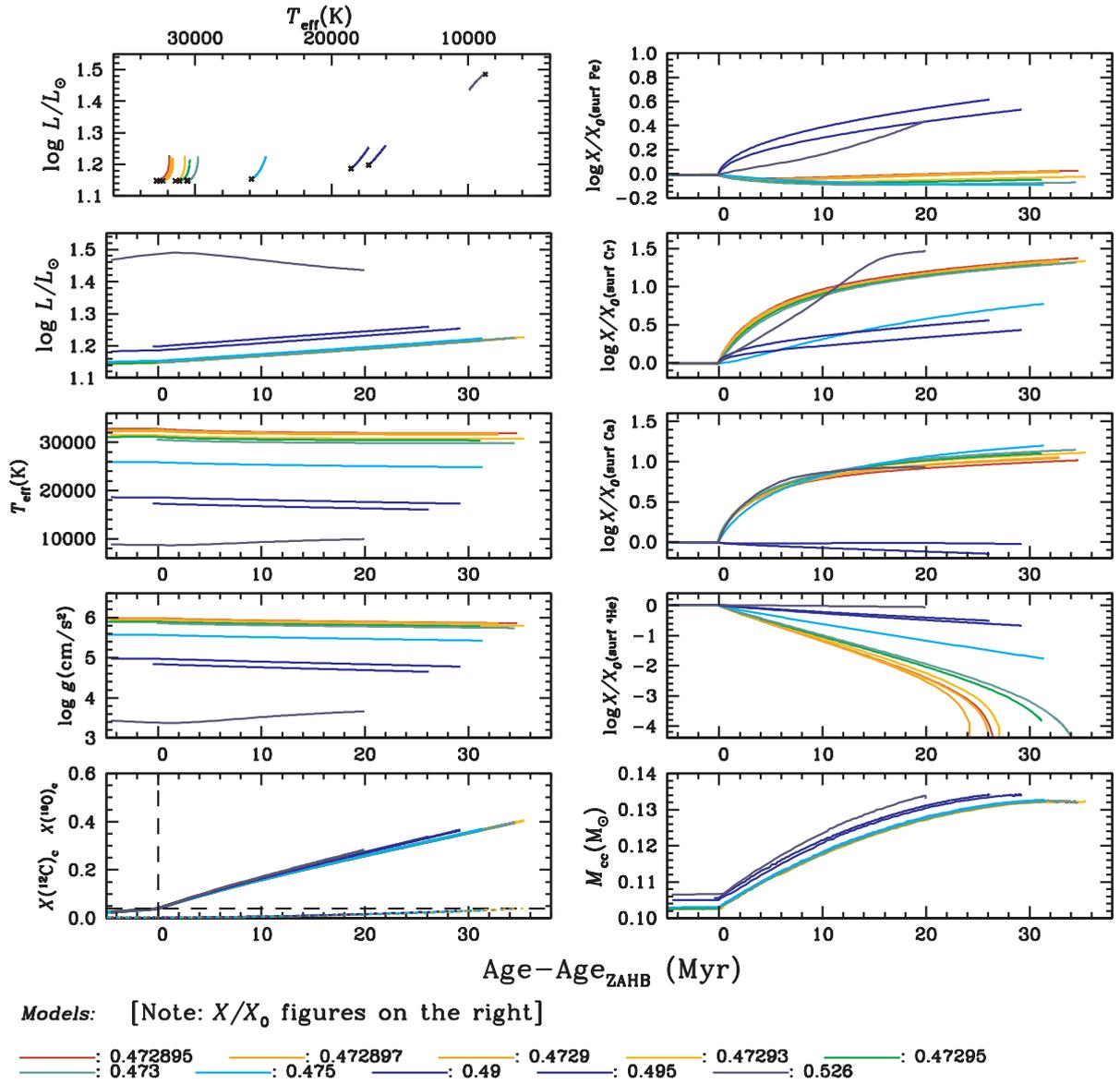}
      \caption{ Models with $Z_0 = 0.02$. See caption of Fig.\,\ref{fig:modelsZ0001}.  }
         \label{fig:modelsZ02}
   \end{figure*}

Some of the properties of all evolutionary models calculated, grouped by metallicity.  All models in Figs.\,\ref{fig:modelsZ0001} to \ref{fig:modelsZ02} were calculated with a turbulent diffusion coefficient given by equations (\ref{eq:DT}) and (\ref{eq:Delta-M-0}) with $\Delta M_0 = 10^{-7.5}$\,\Msol{}. See Sect.\,\ref{sec:Calculations}.

\clearpage
\newpage
\begin{deluxetable}{llrrrcrc}
\tabletypesize{\scriptsize}
\tablecaption{Models illustrated in this article, with their ZAHB characteristics\label{tbl-1}}
\tablewidth{0pt}
\tablehead{
\colhead{$Z_0$\tablenotemark{a}} & \colhead{$M_*$} & \colhead{$L_*$} 
  & \colhead{$T_\mathrm{eff}$} & \colhead{log $g$} 
  & \colhead{Turbulence}  & \colhead{Age\tablenotemark{b}} & \colhead{ID\tablenotemark{c}} \\
  & \colhead{(M$_\odot$)} & \colhead{(L$_\odot$)} & \colhead{(K)} & (cm\,s$^{-2}$) &  & \colhead{(Gy)} 
}

\startdata
0.0001 & 0.507849 & 21.109 & 37\,085  & 6.051 & dM-7.5D10K-4  & 13.196371 & 208783 \\  
0.0001 & 0.50785  & 21.110 & 37\,036  & 6.049 & "             & 13.196353 & 208843 \\  
0.0001 & 0.5079   & 21.128 & 36\,006  & 5.999 & "             & 13.196361 & 208840 \\  
0.0001 & 0.5080   & 21.139 & 35\,160  & 5.958 & "             & 13.196358 & 208842 \\  
0.0001 & 0.5085   & 21.195 & 33\,336  & 5.865 & "             & 13.196334 & 208838 \\  
0.0001 & 0.5090   & 21.267 & 32\,342  & 5.811 & "             & 13.196304 & 208793 \\  
0.0001 & 0.5100   & 21.384 & 31\,010  & 5.736 & "             & 13.196255 & 208783 \\  
0.0001 & 0.5150   & 21.903 & 27\,533  & 5.524 & "             & 13.195965 & 211219 \\  
0.0001 & 0.5200   & 22.376 & 25\,514  & 5.386 & "             & 13.195879 & 208794 \\  
0.0001 & 0.5300   & 23.225 & 22\,815  & 5.184 & "             & 13.195582 & 208790 \\  
0.0001 & 0.5500   & 24.969 & 19\,440  & 4.891 & "             & 13.195211 & 208789 \\  
0.0001 & 0.5700   & 26.981 & 17\,168  & 4.657 & "             & 13.194931 & 208787 \\
0.0001 & 0.5900   & 29.432 & 15\,386  & 4.444 & "             & 13.194738 & 208785 \\
0.0001 & 0.5900   & 29.328 & 15\,388  & 4.445 & T5.0D10K-4d   & 13.194820 & \phantom{0}22180 \\
0.0001 & 0.6100   & 32.344 & 13\,842  & 4.233 & dM-7.5D10K-4  & 13.194621 & 208895 \\
0.0001 & 0.6300   & 35.586 & 12\,442  & 4.021 & "             & 13.194553 & 208847 \\
0.0001 & 0.6500   & 38.906 & 11\,170  & 3.808 & "             & 13.194445 & 209247 \\
0.0001 & 0.6700   & 42.118 & 10\,036  & 3.601 & "             & 13.194436 & 209248 \\
0.0001 & 0.6900   & 45.115 &  9\,050  & 3.404 & "             & 13.194433 & 209251 \\
0.0001 & 0.7300   & 50.354 &  7\,480  & 3.050 & "             & 13.194448 & 209252 \\
\tableline
0.001  & 0.497455 & 19.181 & 36\,542  & 6.058 &  dM-7.5D10K-4 & 12.204222 & 209382 \\
0.001  & 0.49746  & 19.179 & 36\,078  & 6.036 & "             & 12.204233 & 209322 \\
0.001  & 0.4975   & 19.184 & 35\,057  & 5.986 & "             & 12.204219 & 209314 \\
0.001  & 0.4980   & 19.250 & 32\,405  & 5.848 & "             & 12.204192 & 209313 \\
0.001  & 0.4982   & 19.276 & 31\,898  & 5.820 & "             & 12.204193 & 209312 \\
0.001  & 0.4985   & 19.309 & 31\,288  & 5.786 & "             & 12.204252 & 207669 \\
0.001  & 0.4990   & 19.369 & 30\,494  & 5.741 & "             & 12.204227 & 205714 \\
0.001  & 0.5000   & 19.482 & 29\,310  & 5.670 & dM-10.0D10K-4 & 12.204068 & 209839 \\
0.001  & 0.5000   & 19.482 & 29\,318  & 5.671 &  dM-9.0D10K-4 & 12.204068 & 209840 \\
0.001  & 0.5000   & 19.482 & 29\,322  & 5.671 &  dM-8.0D10K-4 & 12.204068 & 209843 \\
0.001  & 0.5000   & 19.478 & 29\,324  & 5.671 &  dM-7.5D10K-4 & 12.204182 & 207709 \\
0.001  & 0.5025   & 19.753 & 27\,377  & 5.548 & "             & 12.203995 & 211226 \\
0.001  & 0.5050   & 19.984 & 26\,007  & 5.456 & "             & 12.203976 & 208085 \\
0.001  & 0.5100   & 20.418 & 24\,005  & 5.312 & "             & 12.203820 & 207668 \\
0.001  & 0.5200   & 21.238 & 21\,284  & 5.094 & "             & 12.203564 & 207704 \\
0.001  & 0.5300   & 22.135 & 19\,343  & 4.918 & "             & 12.203353 & 207706 \\
0.001  & 0.5500   & 24.927 & 16\,403  & 4.596 & "             & 12.203071 & 207707 \\
0.001  & 0.5600   & 27.324 & 15\,043  & 4.414 & "             & 12.202989 & 207467 \\
0.001  & 0.5700   & 30.435 & 13\,627  & 4.203 & "             & 12.202952 & 205899 \\
0.001  & 0.5800   & 33.931 & 12\,169  & 3.967 & "             & 12.202947 & 206117 \\
0.001  & 0.5900   & 37.393 & 10\,734  & 3.714 & "             & 12.202944 & 207665 \\
0.001  & 0.6000   & 40.585 &  9\,387  & 3.453 & "             & 12.202956 & 208081 \\
0.001  & 0.6100   & 43.386 &  8\,166  & 3.189 & "             & 12.202966 & 208083 \\
\tableline
0.004  & 0.48090  & 16.254 & 35\,068  & 6.044 & dM-7.5D10K-4  & 11.072471 & 209535 \\
0.004  & 0.48095  & 16.244 & 33\,257  & 5.952 & "             & 11.072671 & 206918 \\
0.004  & 0.4810   & 16.258 & 32\,667  & 5.921 & "             & 11.072670 & 206852 \\
0.004  & 0.4820   & 16.368 & 29\,282  & 5.728 & "             & 11.072520 & 206850 \\
0.004  & 0.4835   & 16.516 & 27\,311  & 5.605 & "             & 11.072435 & 211224 \\
0.004  & 0.4850   & 16.647 & 26\,021  & 5.519 & "             & 11.072454 & 206849 \\
0.004  & 0.4875   & 16.854 & 24\,447  & 5.407 & "             & 11.072200 & 211225 \\
0.004  & 0.5000   & 17.805 & 19\,923  & 5.039 & "             & 11.071824 & 206917 \\
0.004  & 0.5200   & 20.075 & 15\,842  & 4.606 & "             & 11.071256 & 206832 \\
0.004  & 0.5400   & 28.002 & 11\,683  & 3.949 & "             & 11.071092 & 206914 \\
0.004  & 0.5600   & 36.861 &  7\,265  & 3.020 & "             & 11.071109 & 206960 \\
0.004  & 0.5700   & 39.702 &  5\,793  & 2.602 & "             & 11.071069 & 206962 \\
\tableline
0.02   &0.472895  & 14.060 & 32\,783  & 5.982 &  dM-7.5D10K-4 & 11.712004 & 209540 \\
0.02   &0.472897  & 14.061 & 32\,547  & 5.970 & "             & 11.712016 & 209539 \\
0.02   & 0.47290  & 14.064 & 32\,347  & 5.959 & "             & 11.712347 & 203666 \\
0.02   & 0.47293  & 14.061 & 31\,429  & 5.909 & "             & 11.711976 & 204568 \\
0.02   & 0.47295  & 14.065 & 31\,108  & 5.891 & "             & 11.711949 & 200008 \\
0.02   &  0.4730  & 14.070 & 30\,552  & 5.860 & "             & 11.711964 & 199964 \\
0.02   &  0.4750  & 14.253 & 25\,862  & 5.566 & "             & 11.711779 & 199984 \\
0.02   &  0.4900  & 15.363 & 18\,570  & 4.972 & "             & 11.711468 & 204456 \\
0.02   &  0.4950  & 15.781 & 17\,268  & 4.838 & "             & 11.710998 & 199777 \\
0.02   &  0.5260  & 30.576 &  8\,708  & 3.388 & "             & 11.710788 & 204509 \\
\enddata
\tablenotetext{a}{$Z_0$ is the pre-main-sequence starting metallicity of the models, before 
applying $\alpha$-corrections.  Note that no such correction is applied to $Z_0=0.02$ models.}
\tablenotetext{b}{From zero age \MS.}
\tablenotetext{c}{Identification number which is used to identify models on some of the figures.}
\end{deluxetable}

\end{appendix}

\end{document}